\documentclass[preprint,aps,eqsecnum,amsmath]{revtex4}
\usepackage{amssymb,palatino,float,fancyhdr,mathrsfs,amsbsy,inputenc,bm,dcolumn,rotating}
\usepackage{graphics,graphicx,epsfig}
\begin{document}

\title{Semiclassical features of rotational ground bands}

\author{A. A. Raduta$^{a,b)}$,  R. Budaca $^{a)}$ and Amand Faessler $^{c)}$}

\address{$^{a)}$Institute of Physics and Nuclear Engineering, Bucharest, POB MG6, Romania}

\address{$^{b)}$Academy of Romanian Scientists, 54 Splaiul Independentei, Bucharest 050094, Romania}

\address{$^{c)}$ Instit\" ut f\"ur Theoretische Physik der Universit\"at T\"ubingen, Auf der Morgenstelle 14,
D-72076 T\"ubingen, Germany}

\date{\today}

\begin{abstract}
A time dependent variational principle is used to dequantize a second order quadrupole boson Hamiltonian. The classical equations for the generalized coordinate and the constraint for angular momentum are quantized and then analytically solved. A generalized Holmberg-Lipas formula for energies is obtained.
A similar $J(J+1)$ dependence is provided by the coherent state model (CSM) in the large deformation regime, by using an expansion in powers of $1/x$ for energies, with $x$  denoting a deformation parameter squared. A simple compact expression is also possible  for the near vibrational regime. These three expressions have been used for 44 nuclei covering regions characterized by different dynamic symmetries or in other words belonging to the all known nuclear phases.
Nuclei satisfying the specific symmetries of the critical point in the phase transitions $O(6)\to SU(3)$, $SU(5)\to SU(3)$ have been also considered.
The agreement between the results and the corresponding experimental data is very good. This is reflected in very small  r.m.s. values of deviations. 
\end{abstract}

\pacs{21.10.-k, 21.10.Re, 21.60.Gx, 21.90.+f}

\maketitle

\renewcommand{\theequation}{1.\arabic{equation}}\setcounter{equation}{0}
\label{sec:level1}
\section{Introduction}
One of the big merits of the liquid drop model consists of that it defines in a consistent way the rotational bands. Many theoretical efforts have been made for the description of excitation energies as well as of electromagnetic transitions probabilities.
One of aims, in the beginning era, was to obtain a closed formula for the ground band energies which explain the deviations from the $J(J+1)$ pattern. Various methods have been proposed which were mainly based on the variational moment of inertia principle 
\cite{Maris,Harris,Das}. These approaches proposed for ground band energies a series expansion in terms of $J(J+1)$ term. The weak point of these expansions is that they do not converge for high angular momenta. The first attempt to avoid this difficulty was due Holmberg and Lipas \cite{HoLi} who proposed a square root of a linear expression of $J(J+1)$. This expression proves to work better than a quadratic expression in $J(J+1)$.

Here we address the question whether this formula can be improved such that it could be extended to the region of states with high angular momenta. In the present paper we offer three solutions for this problem, each of them being obtained in a distinct manner. One solution
is based on a semiclassical treatment of a second order quadrupole boson Hamiltonian. The remaining two expressions for the ground band energies are given by asymptotic and near vibrational expansions respectively, of an angular momentum projection formula. 
The three expressions obtained for energies are used for a large number of nuclei. The above sketched project has been achieved according to the following plan. In Section II we present the semiclassical approach in connection with a quadratic quadrupole boson Hamiltonian.
In Section III the angular momentum projection method is described. Numerical applications are presented in Section IV, while the final conclusions are drawn in Section V.

\renewcommand{\theequation}{2.\arabic{equation}}\setcounter{equation}{0}
\section{Semiclassical treatment of a second order quadrupole boson Hamiltonian}
\label{sec:level2}

For a moment we consider the simplest quadrupole  boson ($b^{\dagger}_{2,\mu}, -2\le \mu\le 2$) Hamiltonian:
\begin{equation}
H=A_{1}\sum_{\mu}b_{\mu}^{\dagger}b_{\mu}+A_{2}\sum_{\mu}(b_{\mu}^{\dagger}b_{-\mu}^{\dagger}+b_{\mu}b_{-\mu})(-)^{\mu}.
\end{equation}
Here we are interested to study the classical equations provided by the time dependent variational principle associated to $H$:
\begin{equation}
\delta\int\langle\Psi|H-i\hbar\frac{\partial}{\partial{t}}|\Psi\rangle dt =0.
\label{varpr}
\end{equation}
If the variational states span the whole Hilbert space of boson states, then solving the variational equations is equivalent to solving the time dependent
Schr\"odinger equation which is in general a difficult task. Therefore we restrict the trial function to a coherent state which we hope is the suitable state for describing the semiclassical feature of the chosen system:
\begin{equation}
|\Psi\rangle=Exp\left[z_{0}b_{0}^{\dagger}-z_{0}^{*}b_{0}+z_{2}(b_{2}^{\dagger}+b_{-2}^{\dagger})-z_{2}^{*}(b_{2}+b_{-2})\right].
\end{equation}
Indeed the coherence property results from the obvious equation satisfied by $|\psi\rangle$
\begin{equation}
b_{\mu}|\psi\rangle=(\delta_{\mu0}z_{0}+(\delta_{\mu2}+\delta_{\mu-2})z_{2})|\psi\rangle.
\end{equation}
In order to write explicitly the equations emerging from (\ref{varpr}) we have to calculate first the averages of $H$
\begin{equation}
{\cal H}=\langle \Psi |H|\Psi \rangle,
\end{equation}
as well as of the action operator $-i\hbar \frac{\partial}{\partial{t}}$.
The variational equation (\ref{varpr}) yields the following classical equations for the complex coordinates $z_k$ and $z^*_k$:
\begin{eqnarray}
\frac{\partial{\mathcal{H}}}{\partial{z_{0}}}&=&-i\hbar\dot{z}_{0}^{*},\,\,\,\,\,\frac{\partial{\mathcal{H}}}{\partial{z_{0}^{*}}}=i\hbar\dot{z}_{0},\nonumber\\
\frac{\partial{\mathcal{H}}}{\partial{z_{2}}}&=&-2i\hbar\dot{z}_{2}^{*},\,\,\,\,\,\frac{\partial{\mathcal{H}}}{\partial{z_{2}^{*}}}=2i\hbar\dot{z}_{2}.
\end{eqnarray} 
Note that the coordinates $z_k$ and $z^*_k$ define a classical phase space while ${\cal H}$ plays the role of a classical Hamilton function.
For what follows it is useful to bring these equation to a canonical form. This is achieved by the transformation:
\begin{equation}
q_{i}=2^{(k+2)/4}{\rm{Re}}(z_{k}),\,\,\,\,p_{i}=\hbar2^{(k+2)/4}{\rm{Im}}(z_{k}),\,\,\,\,k=0,2,\,\,\,\,i=\frac{k+2}{2}.
\end{equation}
Indeed, in the new coordinates the classical equations of motion become:
\begin{equation}
\frac{\partial {\cal H}}{\partial q_k}=-\dot{p}_k,\;\;\frac{\partial {\cal H}}{\partial p_k}=\dot{q}_k.
\label{Hameq}
\end{equation}
In terms of the new coordinates, the Hamilton function is written as:
\begin{eqnarray}
\mathcal{H}&=&\frac{A_{1}+2A_{2}}{2}(q_{1}^{2}+q_{2}^{2})+\frac{A_{1}-2A_{2}}{2\hbar^{2}}(p_{1}^{2}+p_{2}^{2})\nonumber\\
&=&\frac{A}{2}(q_{1}^{2}+q_{2}^{2})+\frac{A'}{2\hbar^{2}}(p_{1}^{2}+p_{2}^{2}),
\end{eqnarray}
 where we denoted by $A=A_{1}+2A_{2}$ and $A'=A_{1}-2A_{2}$.
Eqs. (\ref{Hameq}) provide the connection between the generalized momenta and the coordinates time derivatives:
\begin{equation}
p_{1}=\frac{\hbar^{2}\dot{q}_{1}}{A'},\,\,\,\,\,p_{2}=\frac{\hbar^{2}\dot{q}_{2}}{A'}.
\end{equation}
Taking into account these relations, the classical energy function becomes:
\begin{equation}
\mathcal{H}=\frac{\hbar^{2}}{2A'}(\dot{q}_{1}^{2}+\dot{q}_{2}^{2})+\frac{A}{2}(q_{1}^{2}+q_{2}^{2}).
\end{equation}
For what follows it is useful to use the polar coordinates:
\begin{equation}
q_{1}=r\cos{\theta},\,\,\,q_{2}=r\sin{\theta},
\end{equation} 
for the Hamilton function:
\begin{equation}
\mathcal{H}=\frac{\hbar^{2}}{2A'}(\dot{r}^{2}+r^{2}\dot{\theta}^2)+\frac{A}{2}r^{2}.
\end{equation}
The classical system described by ${\cal H}$ is exactly solvable since the number of degrees of freedom is equal to the number of
constants of motion. Indeed, taking the time derivatives of ${\cal H}$ and ${\cal L}_3$, the third component of a pseudo-angular momentum acting in a fictitious boson space, one obtains:
\begin{equation}
\dot{\cal H}=0.\;\;\dot{\cal L}_3=0.
\end{equation} 
The components of the pseudo-angular momentum are defined in Appendix A.
Here we need the conserved component:
\begin{equation}
{\cal L}_3=\frac{1}{2}(q_1p_2-q_2p_1).
\end{equation}
Its constant value is conventionally taken to be:
\begin{equation}
\frac{\hbar^{2}}{2A'}r^{2}\dot{\theta}=L\hbar ,
\end{equation}
which allows us to express the angular variable derivative in terms of the radial one:

\begin{equation}
\dot{\theta}=\frac{2A'L}{\hbar r^{2}}.
\end{equation}
Thus, the energy function written in the reduced space, becomes:
\begin{equation}
\mathcal{H}=\frac{\hbar^{2}}{2A'}\dot{r}^{2}+\frac{2A'L^{2}}{r^{2}}+\frac{A}{2}r^{2}\equiv \frac{\hbar^{2}}{2A'}\dot{r}^{2}+ V_{eff}(r).
\label{classH}
\end{equation}
We recognize in the effective potential energy:
\begin{equation}
V_{eff}(r)=\frac{2A'L^{2}}{r^{2}}+\frac{A}{2}r^{2},
\end{equation}
just the  Davidson potential.

Instead of solving the classical trajectories and then quantizing them, here we first quantize the energy by replacing
\begin{equation}
\frac{\hbar ^2\dot{r}}{A'} \rightarrow -i\hbar \frac{\partial}{\partial r}.
\end{equation}
Thus, one arrives at the Schr\"{o}dinger equation: 
\begin{equation}
\left[-\frac{A'}{2}\frac{\partial^{2}}{\partial{r^{2}}}+\frac{2A'L^{2}}{r^{2}}+\frac{A}{2}r^{2}\right]u(r)=\epsilon u(r).
\end{equation}
Making use of the change of variable and function:
\begin{equation}
x=\sqrt{\frac{A}{A'}}r^{2},\;\;u(r)=e^{-\frac{x}{2}}x^{s}f(x),
\end{equation}
one obtains the following differential equation:
\begin{equation}
\left[x\frac{\partial^{2}}{\partial{x^{2}}}+\left(2s+\frac{1}{2}-x\right)\frac{\partial}{\partial{x}}+\left(\frac{2s^{2}-s-2L^{2}}{2x}+\frac{\epsilon}{2\sqrt{AA'}}-\frac{1}{4}-s\right)\right]f(x)=0.
\end{equation}
This should be compared with the differential equation for the Laguerre polynomials:
\begin{equation}
\left[x\frac{\partial^{2}}{\partial{x^{2}}}+(m+1-x)\frac{\partial}{\partial{x}}+n\right]L_{n}^{m}(x)=0.
\end{equation}
Indeed, the two equations are identical provided the following equations hold:
\begin{equation}
1+m=2s+\frac{1}{2},\,\,\,n=\frac{\epsilon}{2\sqrt{AA'}}-\frac{1}{4}-s,\;\;\; 2s^2-s-2L^2=0.
\label{numcua}
\end{equation}
From the last equation we derive the expression of $s$ as a function of $L$. The positive solution is:
\begin{equation}
s=\frac{1}{4}(1+\sqrt{1+16L^{2}}).
\end{equation}
The second equation (\ref{numcua}) yields for the energy $\epsilon$ the following expression:
\begin{equation}
\epsilon=2\sqrt{(A_{1}^{2}-4A_{2}^{2}})\left(n+\frac{1}{2}+\frac{1}{4}\sqrt{1+16L^{2}}\right),\,\,\,n=0,1,2,...\,\,L=0,1,2,...
\label{energ1}
\end{equation}

An approximative expression may be obtained by expanding first the Davidson potential $V_{eff}$ around its minimum $r_0$ given by the equation:
\begin{equation}
r_{0}^{2}=2L\sqrt{\frac{A'}{A}},
\end{equation}
and truncating the expansion at the quadratic term.
The result for the energy function is:
\begin{equation}
\mathcal{H}=\frac{\hbar^{2}}{2A'}\dot{r}^{2}+2A(r-r_{0})^{2}+2L\sqrt{AA'}.
\end{equation}
Quantizing this Hamilton function we obtain an eigenvalue equation for a harmonic oscillator whose energy is:
\begin{equation}
E_{nL}=2\sqrt{AA'}\left(n+\frac{1}{2}\right)+2L\sqrt{AA'}=2\sqrt{(A_1^2-4a_2^2)}\left(n+\frac{1}{2}+L\right),\;n=0,1,2,...
\label{energ2}
\end{equation}
We remark the fact that the two spectra coincide when $L$ is large:
\begin{equation}
E_{nL}\approx \epsilon_{n,L}, {\rm{for}}~~ L={\rm{large}}.
\end{equation}

Note that the initial boson Hamiltonian could be easily diagonalized by a suitable chosen canonical transformation:
\begin{eqnarray}
\tilde{b}_{\mu}^{\dagger}&=&Ub_{\mu}^{\dagger}-V(-)^{\mu}b_{-\mu},\nonumber\\
\tilde{b}_{\mu}&=&Ub_{\mu}-V(-)^{\mu}b_{-\mu}^{\dagger}.
\end{eqnarray}
Indeed, the coefficients $U$ and $V$ may be chosen such that:
\begin{eqnarray}
\left[\tilde{b}_{\mu},\tilde{b}_{\mu'}^{\dagger}\right]&=&\delta_{\mu\mu'},\nonumber\\
\left[H,\tilde{b}_{\mu}^{\dagger}\right]&=&E\tilde{b}_{\mu}^{\dagger}.
\label{eqforUV}
\end{eqnarray}
The second equation provides a homogeneous system of equations for the tranformation coefficients
\begin{equation}
\left(\begin{array}{c}\,\,A_{1}\,\,\,\,2A_{2}\\
-2A_{2}\,-A_{1}\end{array}\right)\left(\begin{array}{c}U\\V\end{array}\right)=E\left(\begin{array}{c}U\\V\end{array}\right),
\label{rpaeq}
\end{equation}
which determine $U$ and $V$ up to a multiplicative constant which is fixed by the first equation which gives:
\begin{equation}
U^2-V^2=1.
\end{equation}
The compatibility condition for Eq. (\ref{rpaeq}) gives $E=\sqrt{A_{1}^{2}-4A_{2}^{2}}$, and therefore the eigenvalues of $H$ are:
\begin{equation}
E_{n}=\sqrt{A_{1}^{2}-4A_{2}^{2}}\left(n+\frac{5}{2}\right).
\end{equation}
The frequency obtained is half the one obtained through the semiclassical approach. The reason is that here the frequency is associated to each of the 5 degrees of freedom while semiclassically the frequency is characterizing a plane oscillator.
Note that the pseudo-angular momentum $L$ is different from the angular momentum in the laboratory frame describing rotations in the quadrupole boson space:
\begin{equation}
\hat{J}_{\mu}=\sqrt{6}\left(b^{\dagger}_2b_2\right)_{1\mu}.
\end{equation}
The expected value of the angular momentum square is:
\begin{equation}
\langle\Psi|\hat{J}^2|\Psi\rangle =2\left[q_1^2+q_2^2+\frac{1}{\hbar^2}\left(p_1^2+p_2^2\right)\right].
\end{equation}
Since the variational function $|\Psi\rangle$ is not eigenstate of $\hat{J}^2$, the above mentioned average value is not a constant of motion. Indeed, it is easy to check that:
\begin{equation}
\frac{\partial \langle \Psi|\hat{J}^2|\Psi\rangle}{\partial t}=\frac{6}{\hbar^2}\left(A'-A\right)(q_1p_1+q_2p_2)\ne 0.
\end{equation} 
It is instructive to see whether we could crank  the system so that the magnitude of angular momentum is preserved, i.e.
\begin{equation}
\langle\Psi|\hat{J}^2|\Psi\rangle =\hbar^2J(J+1).
\end{equation}
Using the polar coordinates the above equation becomes:
\begin{equation}
\frac{3\hbar^{2}}{A'^{2}}\dot{r}^{2}+\frac{12L^{2}}{r^{2}}+3r^{2}=J(J+1).
\label{classeqJ}
\end{equation}
This equation is treated similarly with the energy equation. Thus by the quantization:
\begin{equation}
\frac{A'^2}{\hbar^2}\dot{r}\rightarrow -\hbar\frac{\partial}{\partial r},
\end{equation}
Eq. (\ref{classeqJ}) becomes a differential equation for the wave function describing the angular momentum:
\begin{equation}
-\frac{\partial^2\Phi}{\partial r^2}+\left(\frac{4L^2}{A'^2r^2}+\frac{r^2}{A'^2}\right)\Phi =\frac{J(J+1)}{3A'^2}\Phi.
\end{equation}
Making the change of variable and function:
\begin{equation}
x=\frac{r^2}{A'},\;\; \Phi =e^{-\frac{x}{2}}x^s\Psi ,
\end{equation}
we obtain the following equation for $\Psi$:
\begin{equation}
x\frac{\partial^2 \Psi}{\partial x^2}+\left(2s-x+\frac{1}{2}\right)\frac{\partial \Psi}{\partial x}+\left(\frac{2s^2-s-\frac{2L^2}{{A'}^2}}
{2x}+\frac{J(J+1)}{12A'}-s-\frac{1}{2}\right)\Psi =0.
\end{equation}
This equation admits the Laguerre polynomials $L^{m'}_{n'}(x)$ with the quantum numbers determined as follows:
\begin{equation}
m'=2s-\frac{1}{2},\;\; s=\frac{1}{4}+\frac{1}{4}\sqrt{1+\frac{16L^2}{{A'}^2}},\;\;\frac{J(J+1)}{12A'}=n'+\frac{1}{2}+\frac{1}{4}
\sqrt{1+\frac{16L^2}{{A'}^2}}.
\label{qnumb}
\end{equation}
The last relation (\ref{qnumb}) can be viewed as an equation determining $L$:
\begin{equation}
L=\left[\left(\frac{J(J+1)}{12}-A'(n'+\frac{1}{2})\right)^2-\left(\frac{A'}{4}\right)^2\right]^{1/2}.
\label{PseudoL1}
\end{equation}
On the other hand taking the harmonic approximation for the potential term in Eq. (\ref{classeqJ}) one obtains the classical equation for a harmonic oscillator from which we get:
\begin{equation}
J(J+1)=12A'\left(n'+\frac{1}{2}\right)+12L.
\end{equation}
Reversing this equation one can express the pseudo-angular momentum $L$ in terms of the angular momentum $J$:
\begin{equation}
L=\frac{J(J+1)}{12}-A'\left(n'+\frac{1}{2}\right).
\label{PseudoL2}
\end{equation}

Replacing, successively, the expressions for $L$,  (\ref{PseudoL1}, \ref{PseudoL2}), into energy equations  (\ref{energ1},  \ref{energ2}), we obtain four distinct expressions for the energies characterizing the starting Hamiltonian $H$.
\begin{eqnarray}
E^{(1)}_{nn'J}&=&\sqrt{AA'}\left[2n+1+\frac{J(J+1)}{6}-A'\left(2n'+1\right)\right],\\
E^{(2)}_{n,n',J}&=&\sqrt{AA'}\left[2n+1+2\sqrt{\left[\frac{J(J+1)}{12}-A'\left(n'+\frac{1}{2}\right)\right]^{2}-\left(\frac{A'}{4}\right)^{2}}\right],\\
E^{(3)}_{nn'J}&=&\sqrt{AA'}\left[2n+1+\frac{1}{2}\sqrt{1+4\left[\frac{J(J+1)}{6}-A'\left(2n'+1\right)\right]^{2}}\right],\\
E^{(4)}_{nn'J}&=&\sqrt{AA'}\left[2n+1+\frac{1}{2}\sqrt{1+4\left[\frac{J(J+1)}{6}-A'\left(2n'+1\right)\right]^{2}-\left(A'\right)^{2}}\right].
\end{eqnarray}
Remark the fact that for a fixed pair of $(n,n')$ each of the above equations define a rotational band: The lowest band corresponds to 
$(n,n')=(0,0)$ and defines the ground band. Except for the band energies $E^{(1)}_{00J}$ which exhibits a $J(J+1)$ pattern the other three bands have the same generic expressions. Thus the excitation energies have the form:
\begin{equation}
E_J=a\left[\sqrt{1+bJ(J+1)+cJ^2(J+1)^2}-1\right].
\label{abc}
\end{equation}
which is  a generalization of the Holmberg-Lipas formula \cite{HoLi}.

We recall that we required that the average value of $\hat{J}^2$ equals $\hbar^2J(J+1)$. Subsequently we eliminated the energy dependence
on the pseudo-angular momentum $L$. In this way we projected approximately the angular momentum from the variational state.
In the next section we shall show that the exact treatment of the angular momentum projection yields also a closed formula for energy as function of $J(J+1)$.
\renewcommand{\theequation}{3.\arabic{equation}}\setcounter{equation}{0}
\label{sec:level3}
\section{The method of angular momentum projected state}

For the sake of simplicity  here we consider a simple form for the variational state
\begin{equation}
|\Psi_g\rangle=e^{d(b^{\dagger}_{20}-b_{20})}|0\rangle
\end{equation}
in connection with the following quadrupole boson Hamiltonian:
\begin{equation}
H=A_1\sum_{\mu}b^{\dagger}_{2\mu}b_{2\mu}+A_2\hat{J}^2.
\end{equation}
The vacumm state for the quadrupole boson operators is denoted by $|0\rangle$ while  $d$ is a real quantity which plays the role of the deformation parameter. The reason is the fact that the average value of the
quadrupole moment, written in the lowest order in terms of quadrupole boson operator, with the function $|\Psi_g\rangle$ is proportional to $d$.
The component of a given angular momentum is obtained by a projection procedure:
\begin{equation}
\varphi^{(g)}_{JM}=N^{(g)}_JP^{J}_{M0}\Psi_g, 
\end{equation}
where $P^J_{MK}$ denotes the angular momentum projection operator:
\begin{equation}
P^J_{MK}=\frac{2J+1}{8\pi^2}\int D^{J*}_{MK}(\Omega)\hat{R}(\Omega)d\Omega,
\end{equation}
with $D^{J*}_{MK}$ denoting the Wigner functions and $\hat{R}(\Omega)$  a rotation defined by the Euler angle $\Omega$.
The system energy  is defined as the average value of $H$ with the projected state:
\begin{equation}
E^{(g)}_J\equiv \langle \varphi^{(g)}_{JM}|H|\varphi^{(g)}_{JM}\rangle =
A_1d^2\frac{I^{(1)}_J(d^2)}{I^{(0)}_{J}(d^2)}+A_2J(J+1),
\end{equation}
where we denoted by $I^{(0)}_{J}$  the overlap integral:
\begin{equation}
I^{(0)}_{J}(x)=2\int_{0}^{1} P_J(y)e^{xP_2(y)}dy,\;\; x=d^2.
\end{equation}
The $k$ derivative of this integral is denoted by:
\begin{equation}
I^{(k)}_{J}(x)=\frac{d^kI^{(0)}_{J}}{dx^k}.
\end{equation}
The normalization constant for the projected state has the expression:
\begin{equation}
\left(N^{(g)}_{J}\right)^{-2}=(2J+1)I^{(0)}_{J}e^{-d^2}.
\end{equation}
These integrals have been analytically calculated in Ref.\cite{Rad82}.
Actually the energies presented here refer to the ground band described by the coherent state model (CSM) which considers simultaneously three interacting bands, ground, beta and gamma.
 In the asymptotic limit of the deformation parameter $d$ the ground band energies have the expression \cite{Rad83}

\begin{equation}
E^{(g,asym)}_J=\frac{A_1}{2}\left[\frac{x-1}{2}+G^{1/2}_{J}\right]+A_2J(J+1),
\label{grsten}
\end{equation}
with
\begin{eqnarray}
G_{J}&=&\frac{9}{4}x(x-2)+\left(J+\frac{1}{2}\right)^2-\frac{4}{9x}\left(3+\frac{10}{x}+\frac{37}{x^2}\right)\nonumber\\
     &+&\frac{2}{3x}\left(1+\frac{10}{3x}+\frac{13}{x^2}\right)J(J+1)-\frac{2}{9x^3}J^2(J+1)^2,\;\;x=d^2.
\end{eqnarray}
It is worth to mention that Eq. (\ref{grsten}) is similar to the generalized HL formula, with the difference that here the coefficients
of the terms $J(J+1)$ and $J^2(J+1)^2$ have explicit expressions in $x$. Moreover there appears an additional $J(J+1)$ term outside the square root symbol.
The expression (\ref{grsten}) is obtained by replacing the series expansion in $1/x$, associated to the ratio $x\frac{I^{(1)}_J}{I^{(0)}_{J}}$,
\begin{eqnarray}
x\frac{I_{J}^{(1)}}{I_{J}^{(0)}}&=&x-1-\frac{1}{3x}-\frac{5}{9x^{2}}-\frac{37}{27x^{3}}+\left(\frac{1}{6x}+\frac{5}{18x^{2}}+\frac{13}{18x^{3}}\right)J(J+1)\nonumber\\
&&-\frac{1}{54x^{3}}j^{2}(J+1)^{2}+\mathcal{O}(x^{-4}),
\label{truncseri}
\end{eqnarray}
by a faster  convergent one.

According to Ref.\cite{Rad84}, for the near vibrational regime ($d$- close to zero) the ground state band energies have the expressions:
\begin{eqnarray}
E^{g,vib}_{J}&=&A_1\left[\frac{J}{2}+\frac{J}{2(2J+3)}x+\frac{9}{2}\frac{(J+1)(J+2)}{(2J+3)^2(2J+5)}x^2\right.\nonumber\\
             &+&\left.\frac{27}{2}\frac{(J+1)(J+2)}{(2J+3)^3(2J+5)(2J+7)}x^3\right] +A_2J(J+1).
\label{g,vib}
\end{eqnarray}
For the sake of completeness we present the derivation of the two expressions for the ground band energies, in the rotational and near vibrational limits, in Appendix B.

\renewcommand{\theequation}{4.\arabic{equation}}\setcounter{equation}{0}
\label{sec:level4}
\section{Numerical results}

Since the expressions (\ref{grsten}), (\ref{truncseri}) and (\ref{g,vib}) are based on series expansion in $1/x$ and $x$, respectively, it is worth showing how far are the truncated expansions from the exact energies. Aiming at this goal in Fig. 1 and Fig. 2 we plotted the
ratio $d^2\frac{I^{(1)}_J}{I^{(0)}_{J}}$ and the associated truncated series for large and small values of $d$ respectively, as functions of $d$ for two angular momenta: $J=12$ and $J=16$. In the case of asymptotic regime we considered also the square root expression.
In this case one defines an existence interval of $d$ for which $G_J\ge 0$. The lower bounds of these intervals for $J$ running from $0$ to $30$ are listed in Table I. From Fig. 1 we see that for $d\ge 3$ the used expressions for energies achieve the convergence even for high angular momenta. Concerning the energies for the near vibrational regime one notes that we use a power series of $x$  and therefore one may think that such an expansion is valid for $x\le 1$. However, we notice that the coefficients of this expansion are depending on $J$ and moreover are under unity numbers. The larger $J$ the smaller are these coefficients. This fact infers that the convergence radius is larger than unity and is an increasing function of the angular momentum. As a matter of fact this is confirmed in the plot shown in Fig. 2.
Comparing the curves from Figs. 1 and 2 one may say that there is a small interval of $d$ were the asymptotic and small $x$ expansions are matched. This allows us to assert that the reunion of the two formulas, (\ref{grsten}) and (\ref{g,vib}),
 assures an overall description of nuclei
ranging from small to large deformation.
In Fig. 3 we plotted the term $G_J$ involved in the energy expression (\ref{grsten}) as a function of the deformation parameter $d$. Except for $J=0$ and $J=2$ all the other functions vanish  for a specific value of $d$ which are, in fact, the lower bounds of the existence interval.

\vspace{0.75cm}
\begin{figure}[hbtp]
\begin{center}
\includegraphics[width=0.80\textwidth]{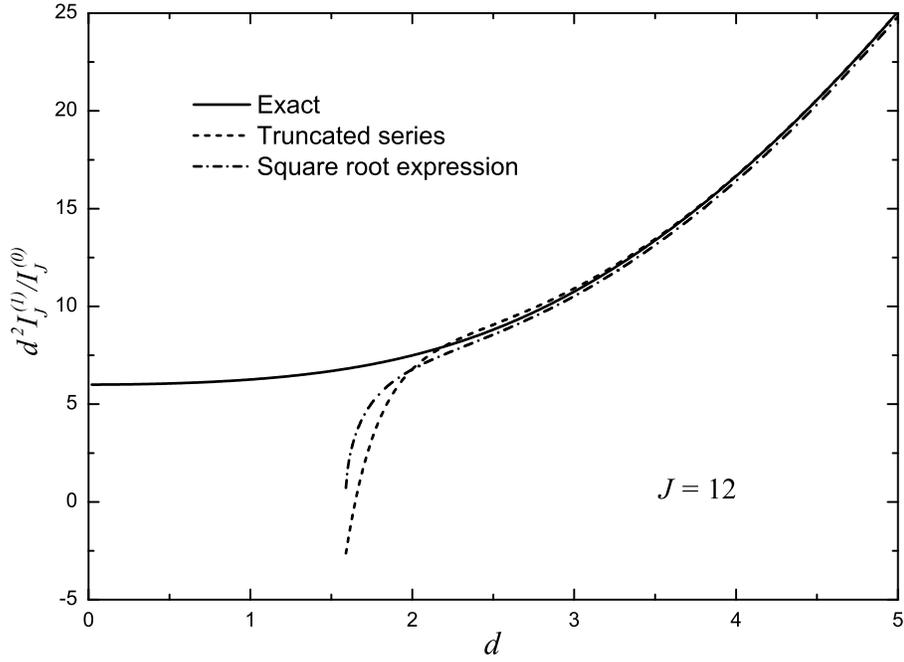}
\includegraphics[width=0.80\textwidth]{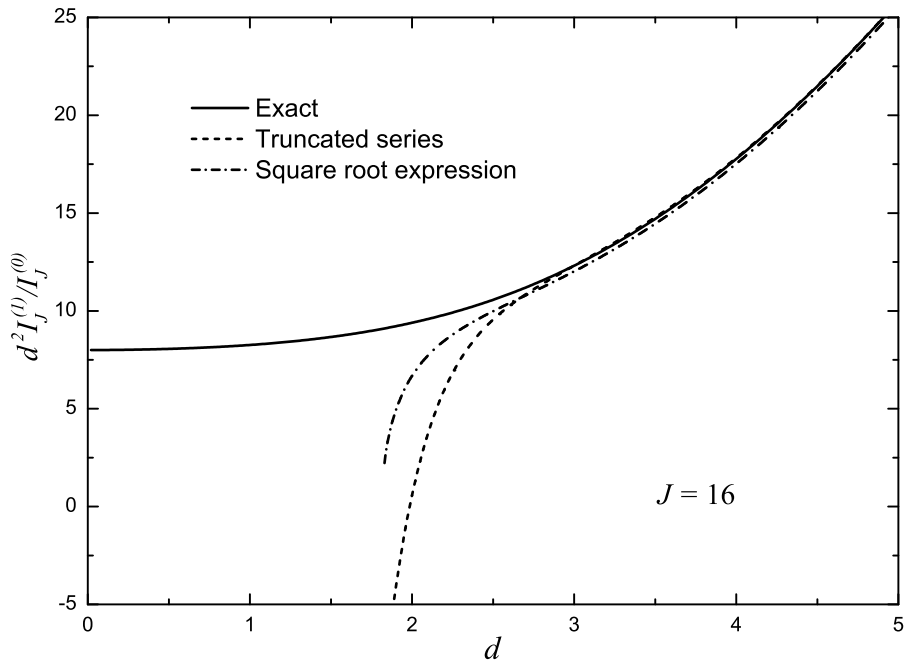}
\end{center}
\caption{ $d^{2}I_{J}^{(1)}/I_{J}^{(0)}$ is plotted as a   function of $d$ for two values of angular momentum. Two approximations of this function are also presented. One is a truncated expansion in $1/x$, while the other one is given by a square 
root expression which is slightly faster convergent than the previously mentioned expansion.}
\end{figure}

\vspace{0.75cm}                                                                                                              \begin{figure}[hbtp]                                                                                                         \begin{center}                                                                                                               \includegraphics[width=0.80\textwidth]{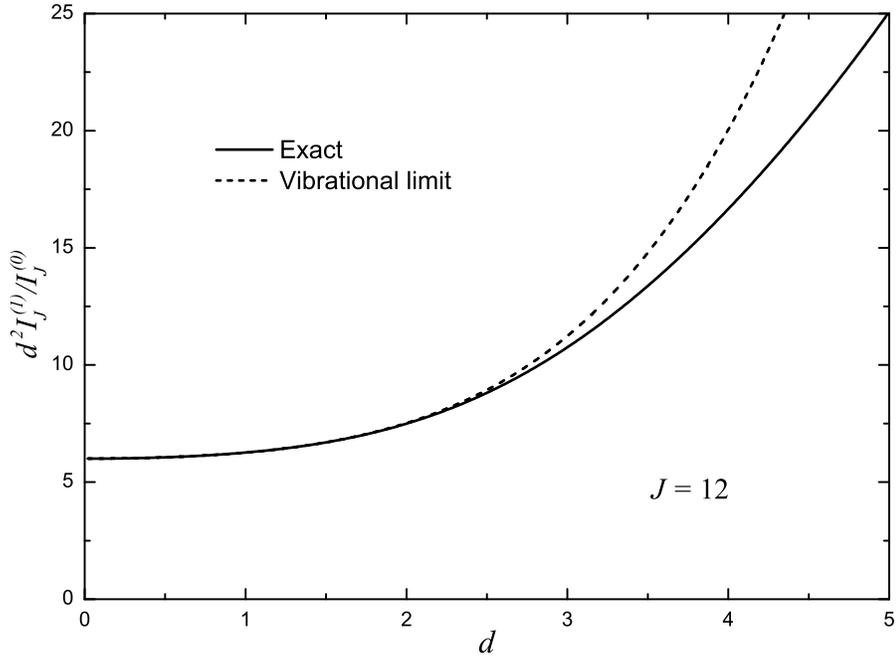}                                                                           \includegraphics[width=0.80\textwidth]{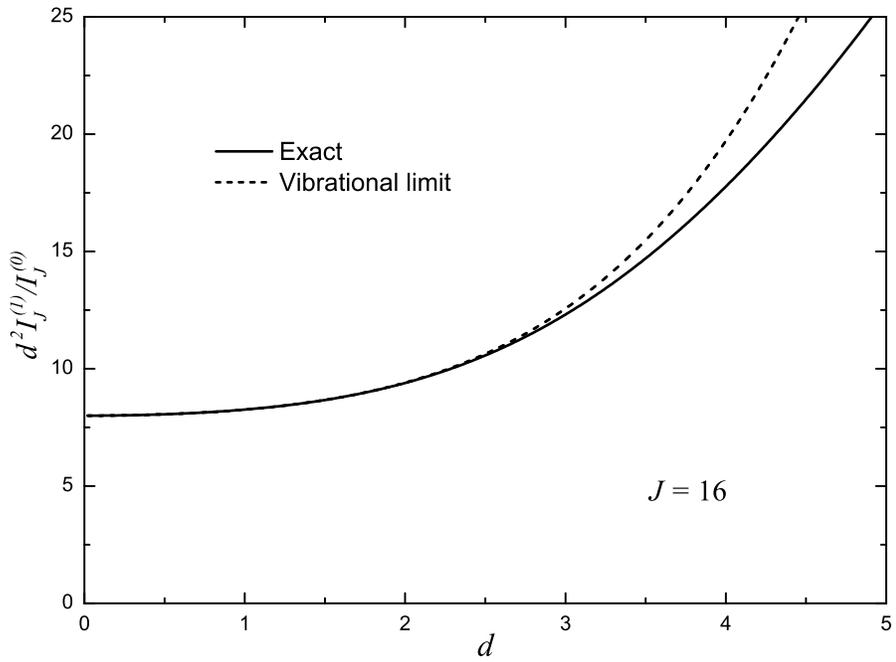}                                                                           \end{center}                                                                                                                 \caption{ $d^{2}I_{J}^{(1)}/I_{J}^{(0)}$ is plotted as a   function of $d$ for  two values of angular momentum. This is compared with the function given by the near vibrational approximation from Eq.(\ref{g,vib}).}                                  
 \end{figure}

\vspace{0.75cm}
\begin{figure}[hbtp]
\begin{center}
\includegraphics[width=0.80\textwidth]{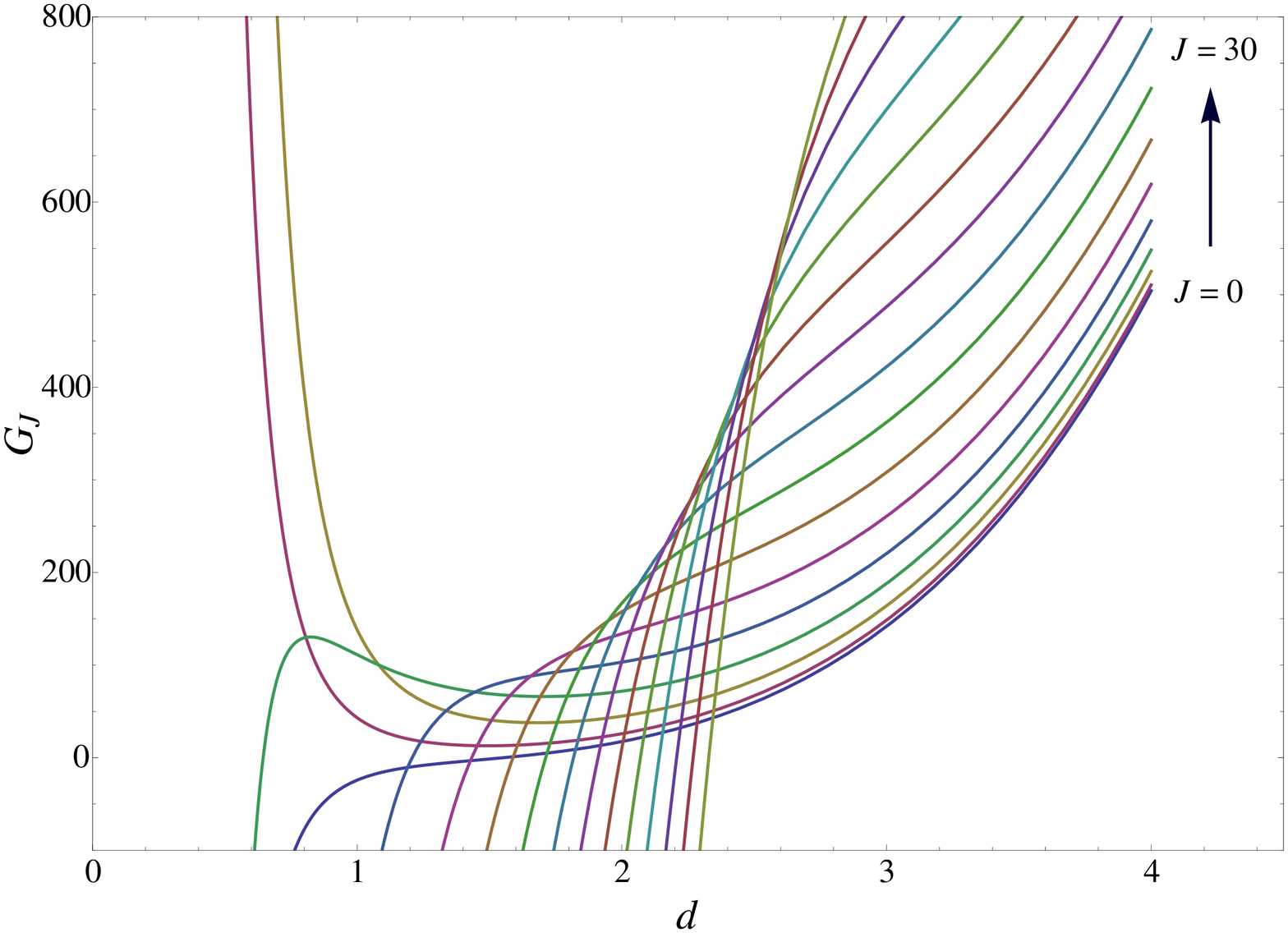}
\end{center}
\caption{$G_{J}$, is plotted as function of $d$ for some angular momenta. Note that excepting the cases of $J=2,4$ all other functions $G_J$ get negative for $d$ smaller than a critical value. These limiting values are listed in Table I.}
\end{figure}
The basic expressions for energies (\ref{abc}), (\ref{grsten}) and (\ref{g,vib}) have been used for a large number of nuclei grouped according to the nuclear phase to which they belong. Thus, for well deformed nuclei behaving like axially deformed rotator the ratio $E_{4^+}/E_{2^+}$ should be close to the value of 3.3 while for the near vibrational region one expects a ratio close to the value 2. Between these two extreme values are placed gamma unstable nuclei where the ratio may run in the interval of 2.5-3.0. The deviation from axial symmetry can affect the ratio mentioned above. Thus $^{228}$Th exhibits some specific feature of a triaxial nucleus with an equilibrium value 
$\gamma^0 = 30^{0}$. The corresponding ratio $E_{4^+}/E_{2^+}$ is equal to 3.24. According to the IBA (Interacting Boson Approximation) model \cite{Ari1,Ari2} the nuclei belonging to the three groups mentioned above are described by the irreducible representations of some dynamic groups as $SU(3)$, $SU(5)$ and $O(6)$. Since the nuclei described by a certain symmetry group exhibit some specific distinct properties one says that these form a certain nuclear phase.
According to Casten \cite{Cast} all nuclei of the periodic table may be placed on the sides of a triangle having in vertexes the three symmetries mentioned above. On each side which links two adjacent symmetries one expects a critical transition point between the two adjacent phases. Few years ago, Iachello \cite{Iache1,Iache2} advanced the idea that each of the critical nuclei laying on the three triangle sides correspond to specific symmetries. Thus, the transition $O(6)\to SU(5)$ is characterized by a critical symmetry which is $E(5)$. Representatives for the  $E(5)$ symmetry are $^{104}$Ru and $^{102}$Pd characterized by specific ratios $E_{4^+}/E_{2^+}=2.48, 2.29 $, respectively. In the transition $SU(5)\to SU(3)$, the critical point
is close to 3. Such nuclei are $^{150}$Nd, $^{152}$Sm, 
$^{154}$Gd and $^{156}$Dy. Indeed, they prove to be critical points for the mentioned phase transition when the entire isotopic chains are considered.

The question  to be answered is whether the compact energy formulas obtained in this paper are able to describe the ground band energies for all nuclei mentioned above.  The theoretical values for energies labeled by Th(1) are obtained with Eq. (\ref{grsten}) if $d$ is large or
with Eq. (\ref{g,vib}) when $d$ is smaller than 2. The calculated energies labeled by Th(2) are obtained with Eq. (\ref{abc}).
The parameters $A_1, A_2, d$ were obtained by a least mean square fitting procedure while $a, b, c$ by fixing the energies of three particular levels.

The agreement of calculated and experimental excitation energies is judged by the r.m.s (root mean square) values of the deviation, denoted by

\begin{equation}
\chi=\sqrt{\sum_{i}^{N}\frac{(E^{Th}_i-E^{Exp}_i)^2}{N}}. 
\end{equation}
The fitting procedure yields for the coefficients $b$ and $c$ double precision numbers, which are presented, in tables, 
in a truncated form. Since the square root formula provides energies which are quite sensitive to small variations for the parameters $b$ and $c$, we give their values with a suitable large number of digits. Indeed, with the listed parameters we get the energies corresponding to the exact parameters yielded by the fitting procedure. Comparing the values of $c$ for 
different nuclei, one remarks that the parameter acquires larger values for smaller deformation parameter. For two nuclei, $^{248}$Cm and 
$^{180}$Os, the parameter $c$ gets negative values which annihilates a part of the contribution coming from the $J(J+1)$ 
term.

\begin{table}[h!]
\begin{center}
{\footnotesize
\begin{tabular}{|c|cccccccccccccccc|}
\hline
$J$&0&2&4&6&8&10&12&14&16&18&20&22&24&26&28&30\\
\hline
$d_{min}$&1.55&0&0&0.65&1.21&1.43&1.59&1.71&1.82&1.91&1.99&2.07&2.14&2.21&2.27&2.33\\
\hline
\end{tabular}}
\end{center}
\caption{The smallest value of $d$, for which $G_{J}$ is positive.}
\label{Table I}
\end{table}
\clearpage

\begin{sidewaystable}[htb!]
\begin{center}
{\scriptsize
\begin{tabular}{|c|c|lll|lll|lll|lll|lll|lll|}
\hline
\multicolumn{2}{|c|}{~}&&$^{228}Th$&&&$^{232}Th$&&&$^{232}U$&&&$^{234}U$&&&$^{236}U$&&&$^{238}U$&\\
\hline
\multicolumn{2}{|c|}{$J^{\pi}$}&~~~Exp.&~~~Th(1)&~~~Th(2)&~~~Exp.&~~~Th(1)&~~~Th(2)&~~~Exp.&~~~Th(1)&~~~Th(2)&~~~Exp.&~~~Th(1)&~~~Th(2)&~~~
Exp.&~~~Th(1)&~~~Th(2)&~~~Exp.&~~~Th(1)&~~~Th(2)\\
\hline
\multicolumn{2}{|c|}{$2^{+}$} &57.76&57.61&57.485  &49.37&49.66&49.31&47.57&47.46
&47.54  &43.50&43.72&43.49   &45.24&45.516&45.22 &44.92&44.73&
44.88\\
\multicolumn{2}{|c|}{$4^{+}$} &186.82 &186.90 &\underline{186.82}&162.12  &162.966 &\underline{162.12}&156.57 
&156.34&156.57  &143.35 & 143.93 &\underline{143.35}&149.48&150.33&\underline{149.48}&148.38 &147.93&
\underline{148.38}\\
\multicolumn{2}{|c|}{$6^{+}$} &378.18 &378.51 &378.78  &333.2   &334.64 &333.63           &322.60   &322.59
&322.96            &296.07 &296.84 &296.07            &309.78 &311.34 &309.97            &307.18  &306.90 &
307.71\\
\multicolumn{2}{|c|}{$8^{+}$} &622.50   &622.73 &623.29            &556.90   &558.25  &557.65           &541.00  &540.94
 &541.33           &497.04  &497.63 &497.02            &522.24  &524.29 &522.724            &518.10   &517.8 &
518.96\\
\multicolumn{2}{|c|}{$10^{+}$}&911.80   &911.49 &\underline{911.80}  &827.0   &827.44 &827.69           &805.80   &805.75
 &805.97           &741.2   &741.3 &741.15            &782.3   &784.32 &783.07             &775.9   &776.26 &
777.492\\
\multicolumn{2}{|c|}{$12^{+}$}&1239.4    &1238.58   &1237.99              &1137.1    &1136.62   &1137.77             &1111.5    &1111.64
  &\underline{1111.5}  &1023.8    &1023.38   &1023.69              &1085.3    &1086.49   &1086.07              &1076.7    &1077.35   &
1078.41\\
\multicolumn{2}{|c|}{$14^{+}$}&1599.5    &1599.27   &1597.63              &1482.8    &1481.12   &1482.77             &1453.7    &1453.81
  &1453.3              &1340.8    &1339.79   &1340.41              &1426.3    &1426.07   &1426.86              &1415.5    &1416.33   
&1416.88\\
\multicolumn{2}{|c|}{$16^{+}$}&1988.1    &1989.81   &\underline{1988.1}   &1858.6    &1857.14   &\underline{1858.6}  &1828.1    &1828.13
  &1827.61             &1687.8    &1687.24   &\underline{1687.8}   &~1800.9    &1798.79   &\underline{1800.9}   &1788.4    &1788.68   &
\underline{1788.4}\\
\multicolumn{2}{|c|}{$18^{+}$}&2407.9    &2407.05   &2407.9               &2262.9    &2261.56   &2262.1              &(2231.5)   &2231.13
  &\underline{2231.5}  &~2063.0    &2062.99   &2063.08              &~2203.9    &2200.85   &2204.11              &2191.1    &2190.24   &
2188.89\\
\multicolumn{2}{|c|}{$20^{+}$}&          &2848.18   &2856.32              &~2691.5    &2691.82   &2690.9              &(2659.7)   &2659.9    &2662.76             &~2464.2    &2464.74   &2464.12              &~2631.7    &2628.97   &2632.91              &2619.1    &2617.29   &
2614.76\\
\multicolumn{2}{|c|}{$22^{+}$}&          &3310.53   &3333.13              &~3144.2    &3145.71   &3143.29             &           &3111.96   &3119.8              &~2889.7    &2890.57   &2889.32              &~3081.2    &3080.31   &3084.22              &3068.1    &3066.53   &
3062.93\\
\multicolumn{2}{|c|}{$24^{+}$}&          &3791.4    &3838.43              &~3619.6    &3621.32   &3618.06             &           &3585.21   &3601.5              &~3339      &3338.8    &3337.54              &(3550)     &3552.43   &3555.4               &3535.3    &3535.05   &
3530.78\\
\multicolumn{2}{|c|}{$26^{+}$}&          &4287.89   &4372.51              &~4116.2    &4116.88   &4114.4              &           &4077.8    &4107.12             &~3808      &3807.96   &\underline{3808.0}   &(4039)     &4043.2    &4044.28              &4018.1    &4020.31   &
4016.08\\
\multicolumn{2}{|c|}{$28^{+}$}&          &4796.81   &4935.76              &(4631.8)   &4630.77   &\underline{4631.8}  &           &4588.11   &4636.2              &(4297)     &4296.74   &4300.17              &(4549)     &4550.77   &\underline{4549.0}   &4517      &4520.08   &
\underline{4517.0}\\
\multicolumn{2}{|c|}{$30^{+}$}&          &5314.41   &5528.64              &(5162)     &5161.41   &5169.96             &           &5114.65   &5188.48             &           &4803.88   &4813.74              &(5077)     &5073.52   &5068.04              &5035      &5032.4    &
5031.98\\
\hline
\multicolumn{2}{|c|}{$\frac{E_{g}^{4^{+}}}{E_{g}^{2^{+}}}$}&3.23&&&3.28&&&3.29&&&3.30&&&3.30&&&3.30&&\\
\multicolumn{2}{|c|}{$\chi$}&&0.56&0.66&&1.13&2.2&&0.17&1.00&&0.53&0.95&&2.21&3.16&&1.40&2.48\\\hline
$A_{1}$&~$a$&182.8720&\multicolumn{2}{c|}{1909.4426}&233.4401&\multicolumn{2}{c|}{3339.3037}&298.6610&\multicolumn{2}{c|}{3472.2796}&221.7839&
\multicolumn{2}{c|}{3443.7877}&386.4548&\multicolumn{2}{c|}{5932.3853}&502.4456&\multicolumn{2}{c|}{5945.5847}\\
$A_{2}$&~$b$&4.0861&\multicolumn{2}{c|}{0.0101534}&3.2430&\multicolumn{2}{c|}{0.00495222}&2.6094&\multicolumn{2}{c|}{0.00458875}&2.9879&
\multicolumn{2}{c|}{0.00423038}&1.9129&\multicolumn{2}{c|}{0.00254985}&1.0667&\multicolumn{2}{c|}{0.00252536}\\
$d$&~$c$&2.7062&\multicolumn{2}{c|}{5.4707523$\cdot10^{-6}$}&3.0822&\multicolumn{2}{c|}{1.026518$\cdot10^{-6}$}&3.3385&
\multicolumn{2}{c|}
{1.102772$\cdot10^{-6}$}&3.2192&\multicolumn{2}{c|}{9.425528$\cdot10^{-7}$}&3.6189&\multicolumn{2}{c|}{7.754542
$\cdot10^{-8}$}&3.8524&
\multicolumn{2}{c|}{6.979548$\cdot10^{-8}$}\\
\hline
\end{tabular}}
\caption{Experimental (Exp.) and theoretical (Th(1) and Th(2)) excitation energies, for several nuclei, $^{228}$Th \cite{Agda1}, 
$^{232}$Th 
\cite{Schmorak1},
 $^{232}$U \cite{Schmorak1}, $^{234}$U \cite{Akovali1}, $^{236}$U \cite{Schmorak2}, $^{238}$U \cite{Chukreev1}, are given in units of keV. The predictions labeled by Th(1) are obtained with the square root formula given by the asymptotic expansion of the CSM-ground band  energies while Th(2) are obtained by the generalized  HL expression. The parameters for Th(1) calculations, i.e. $A_{1},A_{2},d$ were obtained by a least  mean square procedure while those of the set Th(2), $a, b, c$, by fixing three particular energy levels, which are underlined.
The obtained parameters are also listed. To have a hint about the agreement between the theoretical and experimental excitation energies, for each case the r.m.s value of discrepancies, denoted by $\chi$, is also given. The values of $A_{1},A_{2},a,\chi$ are given in keV while,
$d,b,c$ are dimensionless. Having in view a possible classification of the considered nuclei, the ratio ${E^4_g}^+/{E^2_g}^+$ is also 
given.}
\end{center}
\label{Table II}
\end{sidewaystable}
\clearpage

\begin{sidewaystable}[htb!]
\begin{center}
{\scriptsize
\begin{tabular}{|c|c|lll|lll|lll|lll|lll|}
\hline
\multicolumn{2}{|c|}{~}&&$^{236}Pu$&&&$^{238}Pu$&&&$^{240}Pu$&&&$^{242}Pu$&&&$^{248}Cm$&\\
\hline
\multicolumn{2}{|c|}{$J^{\pi}$}&~~~Exp.&~~~Th(1)&~~~Th(2)&~~~Exp.&~~~Th(1)&~~~Th(2)&~~~Exp.&~~~Th(1)&~~~Th(2)&~~~Exp.&~~~Th(1)&~~~Th(2)&~~~Exp.&~~~Th(1)&~~~Th(2)\\
\hline
\multicolumn{2}{|c|}{$2^{+}$} &44.63&44.514&44.602          &44.076&43.818&44.06 &42.824&42.88&42.82 &44.54&44.14&44.49         &43.40&
43.41&43.33\\
\multicolumn{2}{|c|}{$4^{+}$} &147.45 &147.23 &\underline{147.45}&145.95 &145.25 &\underline{145.95}&141.69 &141.84  &\underline{141.69}&
147.3  &146.29 &\underline{147.3}&143.6  &143.80 &\underline{145.6}\\
\multicolumn{2}{|c|}{$6^{+}$} &305.80 &305.55 &305.82           &303.38  &302.44 &303.61 &294.32 &294.52 &249.34 &306.4  &304.52 &
306.20          &298.1  &299.18 &298.88\\
\multicolumn{2}{|c|}{$8^{+}$} &515.7  &515.80 &515.939  &513.58  &512.76 &514.16 &497.52  &497.73 &497.61 &518.1  &516.00  &517.98          &505.0  &506.59 &506.30\\
\multicolumn{2}{|c|}{$10^{+}$}&773.5  &773.6   &\underline{773.5} &773.48  &773.027 &774.231 &(747.8)  &747.93  &747.93 &778.6  &777.19 
&778.86 &760.7  &762.36 &762.29\\
\multicolumn{2}{|c|}{$12^{+}$}&1074.3   &1074.31   &1074.1              &1080.1    &1079.84   &1080.33              &(1041.8)   &1041.61   &1041.71             &1084.4   &1084.14   &1084.78            &1061.3   &1062.39   &1062.77\\
\multicolumn{2}{|c|}{$14^{+}$}&1413.6   &1413.31   &\underline{1413.6}  &1429.1    &1429.73   &\underline{1429.1}   &(1375.6)   &1375.54   &\underline{1375.6}  &1431.7   &1432.67   &\underline{1431.7} &1402.5   &1402.35   &1403.35\\
\multicolumn{2}{|c|}{$16^{+}$}&1786.0   &1786.21   &1788.29             &1818.5    &1819.41   &1817.49              &(1746.9)   &1746.84   &1746.67             &1816.7   &1818.59   &1815.83            &1779.6   &1777.89   &\underline{1779.6}\\
\multicolumn{2}{|c|}{$18^{+}$}&         &2188.99   &2195.01             &2244.9    &2245.79   &2242.86              &(2153.1)   &2153.04   &2152.48             &2236.0   &2237.82   &2233.73            &2187.7   &2184.82   &2187.09\\
\multicolumn{2}{|c|}{$20^{+}$}&         &2618.05   &2631.13             &2705.7    &2706.08   &2703.0               &(2591.9)   &2592.01   &2591.07             &2686     &2686.51   &2682.38            &2621.5   &2619.16   &2621.59\\
\multicolumn{2}{|c|}{$22^{+}$}&         &3070.18   &3094.55             &3198.8    &3197.78   &3196.12              &(3062.2)   &3061.95   &3060.93             &3163     &3161.07   &3159.2             &3077.2   &3077.22   &3079.06\\
\multicolumn{2}{|c|}{$24^{+}$}&         &3552.59   &3583.61             &3720.8    &3718.68   &\underline{3720.8}   &(3560.9)   &3561.32   &\underline{3560.9}  &3662     &3658.24   &\underline{3662.0} &3552.4   &3555.65   &3555.75\\
\multicolumn{2}{|c|}{$26^{+}$}&         &4032.81   &4097.01             &4265.2    &4266.82   &4275.97              &(4089)     &4088.78   &4090.18             &4172     &4175.02   &4188.99            &4048.2   &4051.4    &\underline{4048.2}\\
\multicolumn{2}{|c|}{$28^{+}$}&         &4538.68   &4633.81             &          &4840.48   &4860.82              &           &4643.16   &4648.2              &         &4708.76   &4738.69            &4564.5   &4561.71   &4553.19\\
\multicolumn{2}{|c|}{$30^{+}$}&         &5058.3    &5193.28   &              &5438.15   &5474.79               &           &5223.42   &5234.61             &         &5257.04   &5309.9             &         &5084.14   &5067.79\\
\hline
\multicolumn{2}{|c|}{$\frac{E_{g}^{4^{+}}}{E_{g}^{2^{+}}}$}&3.30&&&3.31&&&3.31&&&3.31&&&3.31&&\\
\multicolumn{2}{|c|}{$\chi$}&&0.18&1.72&&0.99&3.24&&0.19&0.57&&1.89&4.98&&1.95&3.28\\
\hline
$A_{1}$&~$a$~&486.0023&\multicolumn{2}{c|}{4857.3300}&397.4155&\multicolumn{2}{c|}{5443.1075}&232.3635
&\multicolumn{2}{c|}{4487.9012}&692.5499&
\multicolumn{2}{c|}{6375.5552}&705.2236&\multicolumn{2}{c|}{10281.7179}\\
$A_{2}$&~$b$~&1.2356&\multicolumn{2}{c|}{0.00307187}&2.4878&\multicolumn{2}{c|}{0.00270563}&3.42535
&\multicolumn{2}{c|}{0.0031902}&0.3216&
\multicolumn{2}{c|}{0.00233268}&0.0133&\multicolumn{2}{c|}{0.00140828}\\
$d$&~$c$~&3.8514&\multicolumn{2}{c|}{4.909159$\cdot10^{-7}$}&3.9353&\multicolumn{2}{c|}{5.862828$\cdot10^{-7}$}&3.4867&
\multicolumn{2}{c|}{8.397898$\cdot10^{-7}$}&4.2578&\multicolumn{2}{c|}{2.196286$\cdot10^{-7}$}&4.2426
&\multicolumn{2}{c|}{-9.361609$\cdot10^{-8}$}\\
\hline
\end{tabular}}
\caption{The same as in Table II but for a different set of nuclei: $^{236}$Pu \cite{Schmorak2}, $^{238}$Pu 
\cite{Chukreev1}, $^{240}$Pu \cite{Chukreev2}, $^{242}$Pu \cite{Akovali2}, $^{248}$Cm \cite{Akovali3}.}
\label{Table III}
\end{center}
\end{sidewaystable}
\clearpage

\begin{sidewaystable}[htb!]
\begin{center}
{\scriptsize
\begin{tabular}{|c|c|lll|lll|lll|lll|lll|lll|}
\hline
\multicolumn{2}{|c|}{~}&&$^{154}Nd$&&&$^{156}Nd$&&&$^{156}Sm$&&&$^{158}Sm$&&&$^{160}Gd$&&&$^{162}Gd$&\\
\hline
\multicolumn{2}{|c|}{$J^{\pi}$}&~~~Exp.&~~~Th(1)&~~~Th(2)&~~~Exp.&~~~Th(1)&~~~Th(2)&~~~Exp.&~~~Th(1)&~~~Th(2)&~~~Exp.&~~~Th(1)&~~~Th(2)&~~~
Exp.&~~~Th(1)&~~~Th(2)&~~~Exp.&~~~Th(1)&~~~Th(2)\\
\hline
\multicolumn{2}{|c|}{$2^{+}$} &70.8&70.55&70.70 &66.9&67.13&67.06 &75.89&75.60&
75.633 &(72.8)&72.75&72.581&75.26&75.153&75.22&72.1&71.73&71.89\\
\multicolumn{2}{|c|}{$4^{+}$} &233.2 &233.80 &\underline{233.2}&221.8 &221.93 &\underline{221.8}&249.71 &249.64 &
\underline{249.71}&(240.3) &240.538 &\underline{240.3}&248.52 &248.34 &\underline{248.52}&237.3 &236.94 &\underline{237.3}\\
\multicolumn{2}{|c|}{$6^{+}$} &481.9 &481.65 &282.13 &460.4 &460.48 &460.42&517.07 &517.13 &
517.21 &(498.4) &499.16  &499.41  &514.75 &514.77  &514.02  &490.8 &490.93 &491.25\\
\multicolumn{2}{|c|}{$8^{+}$} &810.1 &810.26 &810.48 &777.9 &777.76 &\underline{777.9}&871.9  &871.90 &
\underline{871.9} &(844.5) &843.26 &\underline{844.5}&867.9  &868.0 &868.15 &827.3 &827.32&\underline{827.3}\\
\multicolumn{2}{|c|}{$10^{+}$}&1210.8  &1211.23   &\underline{1210.8} &1168.9  &1168.57   &1168.87 &1307.4   &1307.72   &1307.67
 &(1266.7)  &1267.24   &1269.02            &1300.7   &1300.84   &\underline{1300.7}  &1238.9  &1239.09   &1238.75\\
\multicolumn{2}{|c|}{$12^{+}$}&~1677.3  &1677.3    &1676.12            &1628.4  &1628.05   &1628.23            &1819.3   &1818.97   &
\underline{1819.3}  &(1765.8)  &1765.92   &\underline{1765.8} &1806.3   &1806.12   &1805.6              &1719.5  &1719.26   &
\underline{1719.5}\\
\multicolumn{2}{|c|}{$14^{+}$}&2202.4  &2201.89   &2200.46            &2151.6  &2151.99   &\underline{2151.6} &2400.8   &2400.87   
&2402.81
 &(2334.9)  &2334.77   &2327.52   &2377.3   &2377.21   &2376.51  &2261.3  &2261.39   &2264.46\\
\multicolumn{2}{|c|}{$16^{+}$}&~2779.0  &2779.24   &\underline{2779.0} &2737.0  &2736.83   &2735.47  &  &3049.5    &3055.43
  &          &2969.98   &2947.08            &3008.1   &3008.19   &\underline{3008.1}  &        &2859.77   &2869.72\\
\multicolumn{2}{|c|}{$18^{+}$}&(3399.3)?&3404.46   &3408.08            &        &3379.62   &3377.18            &         &3761.62   &3775.4
              &          &3668.38   &3617.8             &         &3693.96   &3696.07             &        &3509.48   &3532.43\\
\multicolumn{2}{|c|}{$20^{+}$}&         &4073.36   &4085.0             &        &4077.89   &4074.82            &         &4534.54   
&4561.74
   &          &4427.32   &4333.63            &         &4430.11   &4437.09             &        &4206.31   &4250.64\\
\hline
\multicolumn{2}{|c|}{$\frac{E_{g}^{4^{+}}}{E_{g}^{2^{+}}}$}&3.29&&&3.32&&&3.29&&&3.30&&&3.30&&&3.29&&\\
\multicolumn{2}{|c|}{$\chi$}&&0.32&0.82&&0.23&0.57&&0.21&0.78&&0.60&2.95&&0.12&0.40&&0.23&1.21\\
\hline
$A_{1}$&~$a$~&441.8371&\multicolumn{2}{c|}{5240.0423}&241.6860&\multicolumn{2}{c|}{5338.3543}&251.2029&\multicolumn{2}{c|}{4267.15}&259.4635&
\multicolumn{2}{c|}{12111.9787}&485.7106&\multicolumn{2}{c|}{6680.1981}&460.9196&\multicolumn{2}{c|}{4947.3299}\\
$A_{2}$&~$b$~&4.3446&\multicolumn{2}{c|}{0.00451888}&6.5982&\multicolumn{2}{c|}{0.00420209}&7.4572&\multicolumn{2}{c|}{0.00593371}&7.2031&
\multicolumn{2}{c|}{0.00200342}&4.8487&\multicolumn{2}{c|}{0.00376825}&4.5633&\multicolumn{2}{c|}{0.00486469}\\
$d$&~$c$~&3.4184&\multicolumn{2}{c|}{1.524693$\cdot10^{-6}$}&3.24587&\multicolumn{2}{c|}{1.952257$\cdot10^{-6}$}&3.1471&
\multicolumn{2}{c|}{4.471374$\cdot10^{-6}$}&3.2474&\multicolumn{2}{c|}{1.233053$\cdot10^{-8}$}&3.5061&\multicolumn{2}{c|}
{1.059795$\cdot10^{-6}$}&3.4845&\multicolumn{2}{c|}{2.343414$\cdot10^{-6}$}\\
\hline
\end{tabular}}
\caption{The same as in Table II but for a different set of nuclei: $^{154}$Nd 
\cite{Reich1}, $^{156}$Nd \cite{Reich2}, $^{156}$Sm \cite{Reich2}, $^{158}$Sm \cite{Helmer1}, $^{160}$Gd 
\cite{Reich3},
$^{162}$Gd \cite{Helmer2}. The last energy level of $^{154}$Nd is uncertain and thereby it was not involved in the fitting procedure.} 
\label{Table IV}
\end{center}
\end{sidewaystable}
\clearpage

In Tables II and III are given the results for some isotopes of Th, U, Pu and Cm. Except for $^{228}$Th, these isotopes are characterized by large values for the deformation parameter $d$. As we already mentioned, $^{228}$Th has features which are specific to the triaxial nuclei.
We notice the small values for the r.m.s. obtained in these cases. 

In Tables IV, V and VI are studied the deformed nuclei belonging to  the isotopic chains of 
Nd, Sm, Gd, Dy, Er, Yb, Hf, W, Os, respectively. The first six situations
can be viewed as deformed branches of the nuclear phase transition $SU(5)\to SU(3)$ while the last three as the deformed branches of the nuclear phase 
transition $O(6)\to SU(3)$.
The nuclei presented in Table VII, VIII and IX are characterized by small $d$ and moreover they satisfy the $O(6)$ (Table VII), the $SU(5)$ (Table VIII)  and  $X(5)$ (Table IX) 
symmetries, respectively. For all these nuclei, the defining equation (\ref{g,vib}) has been used. One notices that this formula for the near vibrational picture describes the excitation energies better than the generalized HL formula. This is reflected by the relative r.m.s. values.

\begin{sidewaystable}[htb!]
\begin{center}
{\scriptsize
\begin{tabular}{|c|c|lll|lll|lll|lll|lll|lll|}
\hline
\multicolumn{2}{|c|}{~}&&$^{162}Dy$&&&$^{164}Dy$&&&$^{166}Er$&&&$^{172}Yb$&&&$^{174}Yb$&&&$^{176}Hf$&\\
\hline
\multicolumn{2}{|c|}{$J^{\pi}$}&~~~Exp.&~~~Th(1)&~~~Th(2)&~~~Exp.&~~~Th(1)&~~~Th(2)&~~~Exp.&~~~Th(1)&~~~Th(2)&~~~Exp.&~~~Th(1)&~~~Th(2)&~~~
Exp.&~~~Th(1)&~~~Th(2)&~~~Exp.&~~~Th(1)&~~~Th(2)\\
\hline
\multicolumn{2}{|c|}{$2^{+}$} &80.66&80.43&80.59  &73.39&73.36&73.37 &80.58&
81.00&80.54 &78.74&78.68&78.71 &76.47&76.48&76.48 &88.35&88.08
&88.23\\
\multicolumn{2}{|c|}{$4^{+}$} &265.66 &265.25 &\underline{265.66}&242.23 &242.21 &\underline{242.23}&264.99 &
266.16 &\underline{264.99}&260.27  &260.18 &\underline{260.27}&253.12 &253.13 &\underline{253.12}&290.18&289.78
&\underline{290.18}\\
\multicolumn{2}{|c|}{$6^{+}$} &548.52 &548.23 &548.79 &501.32 &501.34 &501.41 &545.45&547.04
&545.56&539.98  &540.01 &540.14&526.03 &525.93 &525.83 &596.82  &596.78
&597.32\\
\multicolumn{2}{|c|}{$8^{+}$} &920.50   &921.14  &921.53  &843.68  &843.68 &843.79 &911.21&912.25
 &911.35 &912.12   &912.15 &912.27 &889.93  &889.26 &888.97 &997.74  &997.97
&998.37\\
\multicolumn{2}{|c|}{$10^{+}$}&1374.80    &1374.98   &\underline{1374.8}   &1261.3    &1261.19   &\underline{1261.3}   &1349.64   &1349.13
 &\underline{1349.64}  &1370.07    &1370.11   &\underline{1370.07} &1336.00      &1336.63   &\underline{1336.0}   &(1481.07)  &1481.13
&\underline{1481.07}\\
\multicolumn{2}{|c|}{$12^{+}$}&1901.3     &1901.0    &1900.04              &1745.9    &1745.74   &1745.76              &1846.6    
&1845.07
 &1847.06              &(1907.48)   &1907.52   &1907.22             &(1861)     &1861.27   &1860.23              &(2034.67)  &2034.29
&2033.65\\
\multicolumn{2}{|c|}{$14^{+}$}&2492       &2491.26   &2489.81              &~2289.6    &2289.65   &2289.52              &2389.4    
&2388.37
  &2390.46              &(2518.7)    &2518.62   &2518.17             &(2457)     &2456.66   &2455.3               &(2646.6)   &2646.49
&2645.6\\
\multicolumn{2}{|c|}{$16^{+}$}&3138       &3138.83   &\underline{3138.0}   &(2886.0)   &2885.99   &2885.77              &(2967.4)   
&2968.66
  &\underline{2967.4}   &(3198.4)    &3198.41   &\underline{3198.4}  &(3117)     &3116.76   &3115.52              &(3308.0)   &3308.11
&\underline{3308.0}\\
\multicolumn{2}{|c|}{$18^{+}$}&3838       &3837.79   &3839.85              &(3528.7)   &3528.66   &\underline{3528.7}   &           
&3576.94
  &3566.42              &            &3942.58   &3944.38             &(3836)     &3836.14   &\underline{3836.0}   &(4010.8)   &4010.9
&4013.56\\
\multicolumn{2}{|c|}{$20^{+}$}&           &4583.15   &4591.79              &(4212.3)   &4212.4    &4213.49              &           
&4205.52   &4177.02              &            &4747.52   &4753.51             &(4610)     &4610.05   &4612.7               &           
&4747.85   &4756.41\\
\multicolumn{2}{|c|}{$22^{+}$}&           &5370.65   &5391.23              &           &4932.68   &4936.22              &           
&4847.84   &4789.64              &            &5610.15   &5623.98             &           &5434.34   &5442.38              &           
&5513.01   &5531.95\\
\multicolumn{2}{|c|}{$24^{+}$}&           &6196.67   &6236.35              &           &5685.62   &5693.78              &           
&5498.26   &5355.47              &            &6527.85   &6554.61             &           &6305.41   &6322.51              &           
&6301.26   &6336.55\\
\hline
\multicolumn{2}{|c|}{$\frac{E_{g}^{4^{+}}}{E_{g}^{2^{+}}}$}&3.29&&&3.30&&&3.29&&&3.31&&&3.31&&&3.28&&\\
\multicolumn{2}{|c|}{$\chi$}&&0.48&1.10&&0.07&0.39&&1.14&0.41&&0.05&0.22&&0.33&1.18&&0.23&1.07\\
\hline
$A_{1}$&~$a$~&530.0578&\multicolumn{2}{c|}{6153.5266}&607.1406&\multicolumn{2}{c|}{7551.6774}&850.8099
&\multicolumn{2}{c|}{9490.2037}
&383.6380&\multicolumn{2}{c|}{6605.21025}&587.6927&\multicolumn{2}{c|}{8756.3360}&753.6198&\multicolumn{2}{c|}{6943.5586}\\
$A_{2}$&~$b$&4.4759&\multicolumn{2}{c|}{0.0043870}&3.1089&\multicolumn{2}{c|}{0.003252137}&0.0142&\multicolumn{2}{c|}
{0.002844946}&6.6422&\multicolumn{2}{c|}{0.00398587}&4.5844&\multicolumn{2}{c|}{0.00292061}&2.2576&\multicolumn{2}{c|}{0.00426081}\\
$d$&~$c$~&3.4126&\multicolumn{2}{c|}{1.171720$\cdot10^{-6}$}&3.5835&\multicolumn{2}{c|}{3.4763788$\cdot10^{-7}$}&3.501835&
\multicolumn{2}{c|}{-6.852260$\cdot10^{-7}$}&3.4108&\multicolumn{2}{c|}{1.605260$\cdot10^{-6}$}&3.7067
&\multicolumn{2}{c|}{5.919085$\cdot10^{-7}$}&3.4441&\multicolumn{2}{c|}{2.818646$\cdot10^{-7}$}\\
\hline
\end{tabular}}
\caption{The same as in Table II but for a different set of nuclei: $^{162}$Dy  \cite{Helmer2}, $^{164}$Dy \cite{Balraj1}, $^{166}$Er
 \cite{Shurshikov}, $^{172}$Yb \cite{Balraj2}, $^{174}$Yb \cite{Browne1}, $^{176}$Hf  \cite{Browne2}.}
\label{Table V}
\end{center}
\end{sidewaystable}
\clearpage

\begin{sidewaystable}[h!]
\begin{center}
{\scriptsize
\begin{tabular}{|c|c|lll|lll|lll|lll|lll|}
\hline
\multicolumn{2}{|c|}{~}&&$^{182}W$&&&$^{186}W$&&&$^{178}Os$&&&$^{180}Os$&&&$^{186}Os$&\\
\hline
\multicolumn{2}{|c|}{$J^{\pi}$}&~~~Exp.&~~~Th(1)&~~~Th(2)&~~~Exp.&~~~Th(1)&~~~Th(2)&~~~Exp.&~~~Th(1)&~~~Th(2)&~~~Exp.&~~~Th(1)&~~~Th(2)&~~~Exp.&~~~Th(1)&~~~Th(2)\\
\hline
\multicolumn{2}{|c|}{$2^{+}$} &100.11&99.55&99.82&122.63&119.61 &121.30&131.6&125.48&129.07 &132.11&130.643&129.82&137.16&133.67&134.65\\
\multicolumn{2}{|c|}{$4^{+}$} &329.43&328.81&\underline{329.43}&396.55&393.36&\underline{396.55}&397.7&388.23&\underline{397.7}&408.62
&409.45&\underline{408.62}&434.09&432.25&\underline{434.09}\\
\multicolumn{2}{|c|}{$6^{+}$} &680.50  &680.92 &681.50&809.25 &809.50&811.02 &761.00  &751.75&762.39&795.08&795.92&796.33 &868.94&870.76
&871.64\\
\multicolumn{2}{|c|}{$8^{+}$} &1144.4&1146.29&1146.07&1349.20&1352.17&\underline{1349.20}&1193.80 &1188.34  &1194.95&1257.44&1256.25  
&\underline{1257.44}&1420.94&1422.74&\underline{1420.94}\\
\multicolumn{2}{|c|}{$10^{+}$}&1711.90&1713.7&\underline{1711.9}&2002.4&2003.71&1998.63&1681.6&1685.41&\underline{1681.6}&1767.57&1766.71  &1766.95             &2067.95  &2065.64  &2061.88\\
\multicolumn{2}{|c|}{$12^{+}$}&2372.3&2371.37&2367.76&(2750.9)&2746.52&\underline{2750.9}&2219.4 &2230.28&2216.81&2308.71&2310.46
&\underline{2308.71} &2781.26&2781.84&\underline{2781.26}\\
\multicolumn{2}{|c|}{$14^{+}$}&(3112.3)&3107.73&3103.24&(3562.4)&3564.28&3601.12&2804.3 &2812.15  &2799.12&2875.0&2874.35&2872.48             &3557.7   &3557.76  &3571.28\\
\multicolumn{2}{|c|}{$16^{+}$}&(3909.2)   &3912.02    &\underline{3909.2}   &          &4442.48  &4546.91 &(3429)&3419.0&\underline{3429.0} &        &3446.79  &3451.67             &         &4382.72  &4427.8\\
\multicolumn{2}{|c|}{$18^{+}$}&(4747.1)?  &4774.46    &4777.95              &          &5368.52  &5587.55              &        &4035.32  &4107.78            &        &4016.34  &4041.91             &         &5247.84  &5349.05\\
\multicolumn{2}{|c|}{$20^{+}$}&           &5686.41    &5703.17              &          &6331.66  &6723.24              &        &4637.75  &4837.1             &        &4570.48  &4640.25             &         &6145.34  &6334.65\\
\multicolumn{2}{|c|}{$22^{+}$}&           &6640.27    &6679.82              &          &7322.68  &7954.72              &        &5183.34  &5618.61            &        &5094.27  &5244.62             &         &7067.97  &7385.08\\
\multicolumn{2}{|c|}{$24^{+}$}&           &7629.42    &7703.93              &          &8333.72  &9282.96              &        &5554.72  &6453.89            &        &5568.28  &5853.54             &         &8008.58  &8501.28\\
\hline
\multicolumn{2}{|c|}{$\frac{E_{g}^{4^{+}}}{E_{g}^{2^{+}}}$}&3.29&&&3.23&&&3.02&&&3.09&&&3.16&&\\
\multicolumn{2}{|c|}{$\chi$}&&2.16&10.84&&2.74&14.73&&8.34&2.33&&1.15&1.39&&1.99&5.79\\
\hline
$A_{1}$&~$a$~&1106.3961&\multicolumn{2}{c|}{9719.4346}&1193.1479&\multicolumn{2}{c|}{3759.9451}&217.4255&
\multicolumn{2}{c|}{1297.2467}&398.5784&\multicolumn{2}{c|}{2299.2478}
&539.1025&\multicolumn{2}{c|}{3350.11}\\
$A_{2}$&~$b$~&1.7794&\multicolumn{2}{c|}{0.00343844}&1.0740&\multicolumn{2}{c|}{0.0108515}&8.5931&\multicolumn{2}{c|}{0.0345814}&5.1661
&\multicolumn{2}{c|}{0.01935326}&6.7316&\multicolumn{2}{c|}{0.0136108}\\
$d$&~$c$~&3.7663&\multicolumn{2}{c|}{4.179487$\cdot10^{-7}$}&3.5059&\multicolumn{2}{c|}{1.256224$\cdot10^{-5}$}&2.1840&
\multicolumn{2}{c|}{3.875729$\cdot10^{-5}$}&2.4119&\multicolumn{2}{c|}{-1.080998$\cdot10^{-7}$}&2.7518&
\multicolumn{2}{c|}{9.3005$\cdot10^{-6}$}\\
\hline
\end{tabular}}
\caption{The same as for Table II but for a different set of nuclei: $^{182}$W \cite{Balraj3}, $^{186}$W
 \cite{Baglin1}, $^{178}$Os \cite{Browne3}, $^{180}$Os \cite{Wu}, $^{186}$Os \cite{Baglin1}. The last energy level of $^{182}$W is uncertain and therefore it was not considered in the fitting procedure. }
\label{Table VI}
\end{center}
\end{sidewaystable}
\clearpage

\begin{sidewaystable}[h!]
\begin{center}
{\scriptsize
\begin{tabular}{|c|c|lll|lll|lll|lll|lll|lll|}
\hline
\multicolumn{2}{|c|}{~}&&$^{170}W$&&&$^{174}Os$&&&$^{178}Os$&&&$^{176}Pt$&&&$^{178}Pt$&&&$^{180}Pt$&\\
\hline
\multicolumn{2}{|c|}{$J^{\pi}$}&~~~Exp.&~~~Th(1)&~~~Th(2)&~~~Exp.&~~~Th(1)&~~~Th(2)&~~~Exp.&~~~Th(1)&~~~Th(2)&~~~Exp.&~~~Th(1)&~~~Th(2)&~~~
Exp.&~~~Th(1)&~~~Th(2)&~~~Exp.&~~~Th(1)&~~~Th(2)\\
\hline
\multicolumn{2}{|c|}{$2^{+}$} &156.72&147.91&151.42 &158.60&152.02&149.25&131.6&119.27 &129.07
          &264.0&233.81&220.91 &(170.1)&149.29&149.75 &153.21 &143.63&138.40\\
\multicolumn{2}{|c|}{$4^{+}$} &462.33&472.55 &\underline{462.33}&435.00&432.32&\underline{435.0}&397.7&404.65&
\underline{397.7}&564.1&549.32&\underline{564.1}&(427.1)&426.55 &\underline{427.1}&410.74&414.33&\underline{410.74}\\
\multicolumn{2}{|c|}{$6^{+}$} &875.53&881.93&877.80&777.63&778.99&796.32 &761.00  &769.78&762.39
          &905.6&918.39&952.62 &(764.6)&779.04&778.67&757.07    &765.22&768.11\\
\multicolumn{2}{|c|}{$8^{+}$} &1363.40 &1356.16  &\underline{1363.4}  &1171.93 &1177.95  &1206.54&1193.8 &1196.31  &1194.95
          &1305.7 &1332.77  &1372.98            &(1177.6) &1194.67  &1189.99            &1181.50     &1185.52  &1190.82\\
\multicolumn{2}{|c|}{$10^{+}$}&1901.5  &1888.66  &1902.46             &1617.5  &1624.47  &1656.71            &1681.6 &1678.19
&\underline{1681.6}
 &1764.8 &1789.52  &1824.74            &(1660.4) &1669.39  &\underline{1660.4} &1674.28     &1671.64  &\underline{1674.28}\\
\multicolumn{2}{|c|}{$12^{+}$}&2464.3  &2476.63  &2488.05             &2113.8  &2116.53  &2144.95            &~2219.4 &2212.86  &2216.81
          &2277.0 &2287.36  &2310.5             &(2207.6) &2201.46  &2192.86            &2229.2      &2222.0   &2219.37\\
\multicolumn{2}{|c|}{$14^{+}$}&(3118.0) &3118.64  &\underline{3118.0}  &2656.3  &2653.1   &2672.04            &2804.3 &2799.02  &2799.12
          &2833.5 &2825.61  &\underline{2833.5} &(2811.9) &2789.99  &2790.85            &2841.5      &2835.83  &2828.63\\
\multicolumn{2}{|c|}{$16^{+}$}&(3815.9) &3813.9   &3792.44             &3239.8  &3233.59  &\underline{3239.8} &(3429)  &3435.95
&\underline{3429.0}
   &3423.8 &3403.88  &3396.93            &(3457.5) &3434.47  &\underline{3457.5} &3504.8      &3512.67  &\underline{3504.8}\\
\multicolumn{2}{|c|}{$18^{+}$}&         &4561.93  &4512.55             &3861.8  &3857.66  &3850.29            &        &4123.2   &4107.78
          &4041.80 &4021.95  &4003.64            &(4107.9) &4134.61  &4195.33            &4252.8      &4252.27  &4250.34\\
\multicolumn{2}{|c|}{$20^{+}$}&         &5362.43  &5279.94             &4524.9  &4525.07  &4505.56 &        &4860.49  &4837.1
          &4690.40 &4679.66  &4656.13            &         &4890.21  &5006.32            &            &5054.43  &5067.32\\
\multicolumn{2}{|c|}{$22^{+}$}&         &6215.19  &6096.38             &5233.0  &5235.69  &5207.47            &        &5647.65  &5618.61
          &5377.0 &5376.91  &5356.52            &         &5701.14  &5892.01            &            &5919.06  &5957.41\\
\multicolumn{2}{|c|}{$24^{+}$}&         &7120.05  &6963.58            &5987.10  &5989.4   &5957.69            &        &6484.53  &6453.89
          &6106.60 &6113.64  &\underline{6106.6} &         &6567.32  &6853.56            &            &6846.07  &6921.93\\
\multicolumn{2}{|c|}{$26^{+}$}&         &8076.93  &7883.16             &6786.1  &6786.12  &6757.65            &        &7371.04  &7344.39
          &6878.6 &6889.78  &6907.87            &         &7488.66  &7891.85            &            &7835.39  &7961.93\\
\multicolumn{2}{|c|}{$28^{+}$}&         &         &                    &7628.4  &7625.81  &7608.61            &        &         &                   &       &         &                   &         &         &                   &            &         &       \\
\multicolumn{2}{|c|}{$30^{+}$}&         &         &                    &8511.6  &8508.41  &\underline{8511.60} &        &         &                   &       &         &                   &         &         &                   &            &         &       \\
\multicolumn{2}{|c|}{$32^{+}$}&         &         &                    &9429.7  &9433.9   &9467.53            &        &         &                   &       &         &                   &         &         &                   &            &         &       \\
\hline
\multicolumn{2}{|c|}{$\frac{E_{g}^{4^{+}}}{E_{g}^{2^{+}}}$}&2.95&&&2.74&&&3.02&&&2.14&&&2.51&&&2.68&&\\
\multicolumn{2}{|c|}{$\chi$}&&8.64&11.98&&4.01&23.78&&7.20&2.33&&17.29&37.15&&17.56&31.75&&6.15&8.80\\
\hline
$A_{1}$&~$a$~&301.1197&
\multicolumn{2}{c|}{1404.2605}&253.6741&\multicolumn{2}{c|}{801.1928}&260.4042&\multicolumn{2}{c|}{1297.25}&277.2380&
\multicolumn{2}{c|}{407.296}&222.5347&\multicolumn{2}{c|}{551.9787}&198.9086&\multicolumn{2}{c|}{761.038}\\
$A_{2}$&~$b$~&6.4525&\multicolumn{2}{c|}{0.03768205}&5.3413&\multicolumn{2}{c|}{0.0673811}&6.1596&\multicolumn{2}{c|}{0.0346}&4.9031
&\multicolumn{2}{c|}{0.2278488}&6.8660&\multicolumn{2}{c|}{0.10072468}&7.7625&\multicolumn{2}{c|}{0.0650988}\\
$d$&~$c$~&1.6390&\multicolumn{2}{c|}{3.305334$\cdot10^{-5}$}&1.5935&\multicolumn{2}{c|}{8.260490$\cdot10^{-5}$}&1.6566&
\multicolumn{2}{c|}{3.88$\cdot10^{-5}$}&1.4494&\multicolumn{2}{c|}{3.279647$\cdot10^{-4}$}
&1.5878&\multicolumn{2}{c|}{3.293427$\cdot10^{-4}$}&1.5869&\multicolumn{2}{c|}{1.71825$\cdot10^{-4}$}\\
\hline
\end{tabular}}
\caption{The same as in Table II but for a different set of nuclei:  $^{170}$W \cite{Baglin2}, $^{174}$Os
\cite{Browne1}, $^{178}$Os \cite{Browne3}, $^{176}$Pt \cite{Browne2}, $^{178}$Pt \cite{Browne3}, $^{180}$Pt \cite{Wu}. Also the predictions Th(1) are obtained with the expression (\ref{g,vib}) corresponding to the expansion characterized by $d$- small.}
\label{Table VII} 
\end{center}
\end{sidewaystable}
\clearpage

\begin{sidewaystable}[h!]
\begin{center}
{\scriptsize
\begin{tabular}{|c|c|lll|lll|lll|lll|}
\hline
\multicolumn{2}{|c|}{~}&&$^{108}Te$&&&$^{150}Sm$&&&$^{152}Gd$&&&$^{154}Dy$&\\
\hline
\multicolumn{2}{|c|}{$J^{\pi}$}&~~~Exp.&~~~Th(1)&~~~Th(2)&~~~Exp.&~~~Th(1)&~~~Th(2)&~~~Exp.&~~~Th(1)&~~~Th(2)&~~~Exp.&~~~Th(1)&~~~Th(2)\\
\hline
\multicolumn{2}{|c|}{$2^{+}$} &625.20&605.01&559.90&333.86&305.14&300.10&344.28&336.40&312.00&334.58&337.10&299.71           \\
\multicolumn{2}{|c|}{$4^{+}$} &1289.00  &1296.67  &\underline{1289.0}&773.24&774.07 &\underline{773.238}&755.40&759.65 &
\underline{755.40}&747.04&756.47&\underline{747.04}\\
\multicolumn{2}{|c|}{$6^{+}$} &2047.9  &2075.43  &2083.78 &1278.75&1300.12&1301.78&1227.38   &1233.74  &1239.98              &1224.08 &1227.93  &1241.06             \\
\multicolumn{2}{|c|}{$8^{+}$} &2945.0  &2936.12  &\underline{2945.0}&1836.87  &1857.81  &1857.43            &1746.78   
&1747.55  &1754.67              &1747.82 &1741.62  &1764.45             \\
\multicolumn{2}{|c|}{$10^{+}$}&3886.2  &3876.53  &3883.14           &2433.00   &2438.50   &\underline{2433.0} &2300.4   
 &2297.09  &\underline{2300.4}   &2304.3  &2293.98  &2315.38             \\
\multicolumn{2}{|c|}{$12^{+}$}&4909.10  &4895.60   &\underline{4909.10}&3048.20   &3038.44  &3027.07            &2883.80
    &2880.59  &2880.62              &2892.60  &2883.42  &2895.80              \\
\multicolumn{2}{|c|}{$14^{+}$}&5980.3  &5992.78  &6032.53           &(3675.70)  &3655.73  &3640.20             &3499.20
    &3497.12  &\underline{3499.20}&3508.60  &3509.09  &\underline{3508.60}  \\
\multicolumn{2}{|c|}{$16^{+}$}&        &7167.72  &7261.50            &4305.90   &4289.29  &4273.71            &4142.70
     &4146.16  &4159.89              &4172.70  &4170.53  &4156.89             \\
\multicolumn{2}{|c|}{$18^{+}$}&        &8420.22  &8602.58           &4929.20   &4938.46  &\underline{4929.20} &        
   &4827.39  &4866.08              &4868.60  &4867.44  &4843.65             \\
\multicolumn{2}{|c|}{$20^{+}$}&        &9750.14  &10061.10             &(5592.8)  &5602.83  &5608.38            &    
       &5540.59  &5620.76              &5589.90  &5599.64  &5571.61             \\
\multicolumn{2}{|c|}{$22^{+}$}&        &11157.40    &11641.10             &          &6282.11  &6312.92            &           &6285.64  &6426.51              &6349.9  &6367.01  &6343.24             \\
\multicolumn{2}{|c|}{$24^{+}$}&        &12641.90    &13346.10             &          &6976.11  &7044.47            &           &7062.43  &7285.57              &7160.70  &7169.45  &\underline{7160.70}  \\
\multicolumn{2}{|c|}{$26^{+}$}&        &14203.70    &15178.60             &          &7684.69  &7804.56            &           &7870.89  &8199.80               &8027.50  &8006.90   &8025.91             \\
\hline
\multicolumn{2}{|c|}{$\frac{E_{g}^{4^{+}}}{E_{g}^{2^{+}}}$}&2.06&&&2.32&&&2.19&&&2.23&&\\
\multicolumn{2}{|c|}{$\chi$}&&15.74&34.41&&16.45&22.54&&4.45&14.00&&9.74&15.61\\
\hline
$A_{1}$&~$a$~&580.6812&\multicolumn{2}{c|}{460.8427}&532.8764&\multicolumn{2}{c|}{635.7440}&409.7560&
\multicolumn{2}{c|}{398.91}&395.7184&\multicolumn{2}{c|}{481.5077}\\
$A_{2}$&~$b$~&9.6296&\multicolumn{2}{c|}{0.6424835}&1.7549&\multicolumn{2}{c|}{0.194008}&3.9253&
\multicolumn{2}{c|}{0.360103}&4.3456&\multicolumn{2}{c|}{0.270578}\\
$d$&~$c$~&1.2510&\multicolumn{2}{c|}{1.419787$\cdot10^{-3}$}&1.5563&\multicolumn{2}{c|}{7.927202$\cdot10^{-5}$}
&1.4306&\multicolumn{2}{c|}
{4.278524$\cdot10^{-4}$}&1.4135&\multicolumn{2}{c|}{2.459865$\cdot10^{-4}$}\\
\hline
\end{tabular}}
\caption{The same as for Table VII but for a different set of nuclei: $^{108}$Te \cite{Blachot1}, $^{150}$Sm
\cite{derMat}, $^{152}$Gd \cite{Agda2}, $^{154}$Dy \cite{Reich1}. }
\label{Table VIII}
\end{center}
\end{sidewaystable}
\clearpage

In Table X we present the results for $^{150}$Nd, $^{152}$Sm and $^{154}$Gd. These nuclei play the role of critical points for the phase transitions
$SU(5)\to SU(3)$ in the respective isotopic chain \cite{I.78,I.79,I.80,I.81}. The first two have been presented also in Table IX where they have been described by a formula obtained for a near vibrational regime. However, one expects that for nuclei close to the critical point the other formula using an asymptotic expansion in terms of $1/x$ works as well. This is actually confirmed by the data presented in Table X for  the first two nuclei.
The isotope $^{154}$Gd is supposed to satisfy the so called $X(5)$ symmetry \cite{I.78}. Our results show that the ground band energies of this critical nucleus is described quite well by the compact formulas (\ref{grsten}), (\ref{g,vib}). 

\begin{table}[h!]
\begin{center}
{\footnotesize
\begin{tabular}{|c|c|lll|lll|lll|}
\hline
\multicolumn{2}{|c|}{~}&&$^{150}Nd$&&&$^{152}Sm$&&&$^{156}Dy$&\\
\hline
\multicolumn{2}{|c|}{$J^{\pi}$}&~~~Exp.&~~~Th(1)&~~~Th(2)&~~~Exp.&~~~Th(1)&~~~Th(2)&~~~Exp.&~~~Th(1)&~~~Th(2)\\
\hline
\multicolumn{2}{|c|}{$2^{+}$} &130.21&122.59&126.16 &121.78&107.02&118.00 &137.77&83.82&128.88\\
\multicolumn{2}{|c|}{$4^{+}$} &381.45&386.14&\underline{381.45}&366.48 &372.03&\underline{366.48} &404.19&398.35  
&\underline{404.19}\\
\multicolumn{2}{|c|}{$6^{+}$} &720.4 &726.86 &722.89 &706.88 &719.04&709.74 &770.44&797.667  &786.22\\
\multicolumn{2}{|c|}{$8^{+}$} &1129.7 &1130.85 &\underline{1129.70}&1125.35  &1131.97  &\underline{1125.35}&1215.61 
&1252.12    &1242.82\\
\multicolumn{2}{|c|}{$10^{+}$}&~1599.00    &1593.52  &1595.26           &1609.23  &1605.59  &1603.21            &1725.02 
&1752.12    &1753.75\\
\multicolumn{2}{|c|}{$12^{+}$}&(2119.00)   &2112.90   &\underline{2119.0}&2148.51  &2137.63  &2140.1             &2285.88 
&2293.59    &2307.61\\
\multicolumn{2}{|c|}{$14^{+}$}&(2682.50) &2687.99  &2702.57           &(2736.01) &2726.97  &\underline{2736.01}&2887.82 
&2874.50     &2898.42\\
\multicolumn{2}{|c|}{$16^{+}$}&         &3318.23  &3348.22           &(3362.0)  &3372.98  &3392.27            &3523.3  
&3493.72    &\underline{3523.30}\\
\multicolumn{2}{|c|}{$18^{+}$}&         &4003.30   &4058.16           &          &4075.29  &4110.63            &4178.10  
&4150.56    &4181.18\\
\multicolumn{2}{|c|}{$20^{+}$}&         &4742.97  &4834.33           &          &4833.64  &4892.87            &4859.00  
&4844.58    &4871.98\\
\multicolumn{2}{|c|}{$22^{+}$}&         &5537.10   &5678.39           &          &5647.87  &5740.62            &5573.00  
&5575.48    &5596.23\\
\multicolumn{2}{|c|}{$24^{+}$}&         &6385.58  &6591.67           &          &6517.87  &6655.29            &6328.70  
&6343.07    &6354.73\\
\multicolumn{2}{|c|}{$26^{+}$}&         &7288.35  &7575.26           &          &7443.56  &7638.09            &7130.30  
&7147.20     &7148.47\\
\multicolumn{2}{|c|}{$28^{+}$}&         &8245.35  &8630.02           &          &8424.87  &8690.01            &7978.50  
&7987.76    &\underline{7978.50}\\
\multicolumn{2}{|c|}{$30^{+}$}&         &9256.53  &9756.66           &          &9461.76  &9811.89            &8875.90  
&8864.66    &8845.86\\
\hline
\multicolumn{2}{|c|}{$\frac{E_{g}^{4^{+}}}{E_{g}^{2^{+}}}$}&2.93&&&3.01&&&2.93&&\\
\multicolumn{2}{|c|}{$\chi$}&&5.61&7.92&&9.84&11.43&&23.86&18.45\\
\hline
$A_{1}$&~$a$~&215.8125&\multicolumn{2}{c|}{2867.41}&221.5088&\multicolumn{2}{c|}{1187.15}&348.6711
&\multicolumn{2}{c|}{1913.4867}\\
$A_{2}$&~$b$~&6.7513&\multicolumn{2}{c|}{0.0513}&6.9223&\multicolumn{2}{c|}{0.0344}&4.4983&\multicolumn{2}{c|}{0.0231439}\\
$d$&~$c$~&1.6321&\multicolumn{2}{c|}{1.17$\cdot10^{-4}$}&1.6640&\multicolumn{2}{c|}{6.11$\cdot10^{-5}$}&1.7089
&\multicolumn{2}{c|}{1.051357$\cdot10^{-5}$}\\
\hline
\end{tabular}}
\caption{The same as in Table VII but for a different set of nuclei:
$^{150}$Nd \cite{derMat}, $^{152}$Sm 
\cite{Agda2},
$^{156}$Dy \cite{Reich2}.}
\label{Table IX}
\end{center}
\end{table}

\clearpage

\begin{table}[h!]
\begin{center}
{\footnotesize
\begin{tabular}{|c|c|lll|lll|lll|}
\hline
\multicolumn{2}{|c|}{~}&&$^{150}Nd$&&&$^{152}Sm$&&&$^{154}Gd$&\\
\hline
\multicolumn{2}{|c|}{$J^{\pi}$}&~~~Exp.&~~~Th(1)&~~~Th(2)&~~~Exp.&~~~Th(1)&~~~Th(2)&~~~Exp.&~~~Th(1)&~~~Th(2)\\
\hline
\multicolumn{2}{|c|}{$2^{+}$} &130.21&123.45&126.16   &121.78&119.14&118.00 &123.07&108.23 &117.32     \\
\multicolumn{2}{|c|}{$4^{+}$} &381.45&374.67&\underline{381.45}&366.48&364.57&\underline{366.48}&370.99 &346.04
&\underline{370.99}\\
\multicolumn{2}{|c|}{$6^{+}$} &720.40 &717.10&722.89 &706.88  &704.82&709.74 &717.65 &689.42&728.47 \\
\multicolumn{2}{|c|}{$8^{+}$} &1129.70  &1130.89  &\underline{1129.70}  &1125.35   &1123.09  &\underline{1125.35} &1144.43   &1117.23  &1162.21              \\
\multicolumn{2}{|c|}{$10^{+}$}&1599.00    &1603.40   &1595.26             &1609.23   &1608.64  &1603.21             &1637.04   &1614.03  &1654.28              \\
\multicolumn{2}{|c|}{$12^{+}$}&(2119.00)   &2123.44  &\underline{2119.00}  &2148.51    &2151.71  &2140.10           &2184.67   &2168.74  &2194.41              \\
\multicolumn{2}{|c|}{$14^{+}$}&(2682.5) &2677.98  & 2702.57            &(2736.01)  &2740.64  &\underline{2736.01} &2777.30   &2772.90   &\underline{2777.30}   \\
\multicolumn{2}{|c|}{$16^{+}$}&         &3248.12  & 3348.22            &(3362.0)   &3357.92  &3392.27             &3404.44   &3419.34  &3400.53              \\
\multicolumn{2}{|c|}{$18^{+}$}&         &3798.63  & 4058.16            &           &3968.95  &4110.63             &4087.10    &4101.33  &4063.38              \\
\multicolumn{2}{|c|}{$20^{+}$}&         &4221.54  & 4834.33            &           &4444.16  &4892.87             &4782.30    &4811.93  &4766.07              \\
\multicolumn{2}{|c|}{$22^{+}$}&         &         & 5678.39            &           &         &5740.62             &5519.50    &5543.34  &5509.31              \\
\multicolumn{2}{|c|}{$24^{+}$}&         &         & 6591.67            &           &         &6655.29             &6294.10    &6286.21  &\underline{6294.10}   \\
\multicolumn{2}{|c|}{$26^{+}$}&         &         & 7575.26            &           &         &7638.09             &7055.50    &7028.36  &7121.53              \\
\hline
\multicolumn{2}{|c|}{$\frac{E_{g}^{4^{+}}}{E_{g}^{2^{+}}}$}&2.93&&&3.01&&&3.01&&\\
\multicolumn{2}{|c|}{$\chi$}&&4.83&7.92&&2.93&11.43&&21.22&21.77\\
\hline
$A_{1}$&~$a$&151.3419&\multicolumn{2}{c|}{867.4073}&126.4776&\multicolumn{2}{c|}{1187.1506}&274.9237
&\multicolumn{2}{c|}{1950.92}\\
$A_{2}$&~$b$~&9.5423&\multicolumn{2}{c|}{0.0513027}&10.3917&\multicolumn{2}{c|}{0.034413709}&7.4351
&\multicolumn{2}{c|}{0.02057356}\\
$d$&~$c$~&2.0284&\multicolumn{2}{c|}{1.171278$\cdot10^{-4}$}&2.0136&\multicolumn{2}{c|}{6.109010$\cdot10^{-5}$}&2.4767
&\multicolumn{2}{c|}{1.254659$\cdot10^{-5}$}\\
\hline
\end{tabular}}
\caption{The same as in Table II but for a different set of nuclei: $^{150}$Nd \cite{derMat}, $^{152}$Sm 
\cite{Agda2},
$^{154}$Gd \cite{Reich1}. }
\label{Table X}
\end{center}
\end{table}

\newpage
In the last table (Table XI) we present two nuclei which satisfy the symmetry $E(5)$. These are described with the close formulas
(\ref{g,vib}) and (\ref{abc}). We remark that also in this case the r.m.s. values are small. Experimental data for these nuclei were considered  up to the angular momentum were the first backbending is showing up.

\begin{table}[h!]
\begin{center}
{\footnotesize
\begin{tabular}{|c|c|lll|lll|}
\hline
\multicolumn{2}{|c|}{~}&&$^{104}Ru$&&&$^{102}Pd$&\\
\hline
\multicolumn{2}{|c|}{$J^{\pi}$}&~~~Exp.&~~~Th(1)&~~~Th(2)&~~~Exp.&~~~Th(1)&~~~Th(2)\\
\hline
\multicolumn{2}{|c|}{$2^{+}$} &358.03&348.13&\underline{358.03}&556.43&563.02&520.03\\
\multicolumn{2}{|c|}{$4^{+}$} &888.49&901.59&910.87 &1275.87 &1280.12  &\underline{1275.87}\\
\multicolumn{2}{|c|}{$6^{+}$} &1556.30  &1561.29  &\underline{1556.30}  &2111.35 &2100.12  &2114.71\\
\multicolumn{2}{|c|}{$8^{+}$} &2320.30  &2303.22  &2287.30              &3013.06 &3006.50   &3019.06\\
\multicolumn{2}{|c|}{$10^{+}$}&3111.80  &3119.23  &\underline{3111.80}  &3992.71 &3993.31  &\underline{3992.71}\\
\multicolumn{2}{|c|}{$12^{+}$}&        &4005.75  &4039.35             &5055.10 &5057.87  &5043.55\\
\multicolumn{2}{|c|}{$14^{+}$}&        &4960.97  &5078.09             &6179.80  &6198.80   &\underline{6179.80}\\
\multicolumn{2}{|c|}{$16^{+}$}&        &5983.88  &6234.53             &7428.80  &7415.32  &7408.97\\
\multicolumn{2}{|c|}{$18^{+}$}&        &7073.86  &7513.64             &        &8706.93  &8737.58\\
\multicolumn{2}{|c|}{$20^{+}$}&        &8230.5   &8919.18             &        &10073.30  &10171.10\\
\hline
\multicolumn{2}{|c|}{$\frac{E_{g}^{4^{+}}}{E_{g}^{2^{+}}}$}&2.48&&&2.29&&\\
\multicolumn{2}{|c|}{$\chi$}&&11.33&17.84&&9.88&15.41\\
\hline
$A_{1}$&~$a$~&522.4864&\multicolumn{2}{c|}{562.6938}&649.8648&\multicolumn{2}{c|}{710.6779}\\
$A_{2}$&~$b$~&8.1960&\multicolumn{2}{c|}{0.2738559}&9.2391&\multicolumn{2}{c|}{0.329923}\\
$d$&~$c$~&1.5466&\multicolumn{2}{c|}{9.51995$\cdot10^{-4}$}&1.4174&\multicolumn{2}{c|}{5.317905$\cdot10^{-4}$}\\
\hline
\end{tabular}}
\caption{The same as in Table VII but for a different set of nuclei:$^{104}$Ru \cite{Blachot2}, $^{102}$Pd
\cite{DeFrenne}. These nuclei obey the E(5) symmetry.}
\label{Table XI}
\end{center}
\end{table}
\clearpage

\renewcommand{\theequation}{5.\arabic{equation}}\setcounter{equation}{0}
\label{sec:level5}
\section{Conclusions}

In this Section we summarize the main results obtained in this paper. By a dequantization procedure we associated to a quantum mechanical Hamiltonian which is quadratic in the quadrupole bosons, a time dependent classical equation. The classical Hamiltonian has a separated form, i.e. is a sum of a kinetic  and a potential energy terms. The later one is not depending on momenta and is of Davidson type. We may say that our procedure proves the classical origin of the Davidson potential. The centrifugal term is determined by a pseudo-angular momentum associated to the intrinsic coordinates.
It is worth  mentioning that the constraint for  the angular momentum in the laboratory frame yields a differential equation which is connected to that one corresponding to the energy conservation which results in obtaining a specific angular momentum dependence for  the quantal  energy.
Actually the expression obtained generalizes the Holmberg-Lipas formula, involving under the square root symbol a $J^2(J+1)^2$ term.

A similar expression was obtained by one author (A. A. R.) within the coherent state model (CSM) for a large deformation regime. 
Another compact expression was proposed by CSM for the near vibrational regime, i.e. small nuclear deformation. One of the 
targets of this paper was to prove
that the two compact expressions provided by CSM are able to describe the ground state energies for deformed, near vibrational and transitional nuclei.  
By matching the two expressions one obtains a unitary description for nuclei satisfying different symmetries or, with other words, belonging to various nuclear phases. Similar goal is touched by using a square root formula, a generalization of the Holmberg-Lipas formula, obtained on the base of a semiclassical description.

These descriptions are used for a large number of nuclei (44). The agreement between results and experimental excitation energies is very impressive.
The agreement quality is judged by the small  r.m.s. values of discrepancies.

As a final conclusion we may say that the CSM procedure is able to describe in a realistic fashion the ground state energies for nuclei of different
nuclear phases. An alternative description is given by a square root formula derived as approximate eigenvalue of a quadratic Hamiltonian in quadrupole bosons subject to a constraint due to the angular momentum conservation.

{\bf Acknowledgment.} This work was supported by the Romanian Ministry for Education Research Youth and Sport through the projects
ID-33/2007 and ID-946/2009.

\renewcommand{\theequation}{A.\arabic{equation}}
\setcounter{equation}{0}

\section{Appendix A}

Using the equations of motion for the conjugate variables, one can prove that

\begin{equation}
\dot{\cal L}_3=0~~,~~\dot{\cal H}_1=0,
\label{cmot}
\end{equation}
where ${\cal L}_{3}$ is defined by the following expression:

\begin{equation}
{\cal L}_{3} \equiv \frac{\hbar}{2}
(q_{1} p_{2} - q_{2} p_{1}) =
\frac{\hbar^2}{A'} r^2 \dot{\theta},
\label{L3}
\end{equation}
and has the significance of the third component of the angular momentum defined in the
 phase space, spanned
by the coordinates ($q_{1},p_{1},q_{2},p_{2}$).
The other two components are:
\begin{equation}
{\cal L}_{1} = \frac{\hbar}{4}
((q_{1}^2 + p_{1}^{2} - q_{2}^{2}- p_{2}^{2}),~~
{\cal L}_{2}=\frac{\hbar}{2}(q_{1} q_{2} + p_{1}p_{2}) .
\label{L1andL2}
\end{equation}
Indeed, one easily check that
\begin{equation}
\{{\cal L}_{i},{\cal L}_{k}\}=\hbar\epsilon_{ikj}{\cal L}_{l},
\label{L1L2L3}
\end{equation}
where $\{,\}$ denotes the Poisson bracket while $\epsilon_{ikj}$
the antisymmetric unit tensor.
In virtue of Eq. (\ref{L1L2L3}) the set of functions ${\cal L}_{k}$ with the
Poisson brackets as multiplication
operation, form a classical  $SU_c(2)$ algebra. Moreover, they could be obtained
by averaging with  $\mid\Psi\rangle$, the generators ${\hat L}_k$
\begin{equation}
 {\cal L}_{k} = \langle \Psi \mid {\hat L}_{k} \mid \Psi \rangle; k=1,2,3,
\label{Lboz}
\end{equation}
 of a boson $SU_b(2)$
algebra
defined with the boson operators $b_{0}^{\dagger},b_{\pm 2}^{\dagger}$, as:
\begin{eqnarray}
 {\hat L}_{1}& = &\frac{\hbar}{4}\left[2 b_{0}^{\dagger}b_{0} - ( b_{2}^{\dagger}
+
b_{-2}^{\dagger} )( b_{2} + b_{-2} )\right],\nonumber \\
{\hat L}_{2}& = &\frac{\hbar}{2 \sqrt{2}}\left[b_{0}^{\dagger}(b_{2} + b_{-2})+ (b_{2}^{\dagger} + b_{-2}^{\dagger})b_{0}\right],
\nonumber\\
{\hat L}_{3}&=&\frac{\hbar}{2 \sqrt{2} i}\left[b_{0}^{\dagger}(b_{2} + b_{-2})-
 (b_{2}^{\dagger} + b_{-2}^{\dagger})b_{0}\right].
\label{Lbozi1}
\end{eqnarray}
The equation (2.15) and the correspondence between commutators and Poisson brackets
$[,]  \rightarrow \frac{1}{i}\{,\},$
define a homeomorphism of the boson and classical algebras generated by
$ \{ {\hat L}_{k} \}_{k=1,2,3}$  and $\{{\cal L}_{k}\}_{k=1,2,3}$ respectively.
Note that the boson $SU_b(2)$ algebra does not describe the rotations
in the real configuration space but in a fictitious space.
The conservation law expressed by (\ref{cmot}) is determined by the invariance against
rotation around the 3-rd axis in the fictitious space mentioned above:
$[H,\hat{L}_{3}] = 0$ .
Since the classical system is characterized by two degrees of freedom and, on
the other hand, there are two constants of motion
\begin{equation}
{\cal H} = E~~ ,~~{\cal L}_{3} = L ,
\label{H1eqE}
\end{equation}
the equations of motion are exactly solvable.

\renewcommand{\theequation}{B.\arabic{equation}}
\setcounter{equation}{0}
\section{Appendix B}

By direct calculations we can check that the overlap integral $I^{(0)}_J$ and its first and second derivatives satisfy the following differential equation:
\begin{gather}
\frac{d^{2}I_{J}^{(0)}}{dx^{2}}-\frac{x-3}{2x}\frac{dI_{J}^{(0)}}{dx}-\frac{2x^{2}+J(J+1)}{4x^{2}}I_{J}^{(0)}=0,\,\,\,\,(x=d^{2}).
\end{gather}
By a suitable change of function this equation can be brought to the differential equation characterizing the hypergeometric function of the first rank. Thus the final result for $I^{(0)}_J$ is:
\begin{gather}
I_{J}^{(0)}=\frac{(J!)^{2}}{\left(\frac{J}{2}\right)!(2J+1)!}(6d^{2})^{\frac{J}{2}}e^{-\frac{d^{2}}{2}}{_{1}F_{1}}\left(\frac{1}{2}(J+1),J+\frac{3}{2};\frac{3}{2}d^{2}\right).
\end{gather}
This expression is further used for describing both the asymptotic and vibrational behavior for the excitation energies in the ground band.
Indeed, in the asymptotic region of $d$, the hypergeometric function behaves like:
\begin{gather}                                                                                                                             _{1}F_{1}(a,c;z)=\frac{\Gamma(c)}{\Gamma(a)}e^{z}z^{a-c}[1+\mathcal{O}(|z|^{-1})].                                                                 \end{gather} 
Due to this expression the dominant term of $I^{(0)}_J$ is:
\begin{gather}
I_{J}^{(0)}\sim\frac{e^{x}}{3x}.
\label{I0asim}
\end{gather}
This expression suggests for  $I^{(0)}_J$ , in the asymptotic region, the following form: 
\begin{gather}
I_{J}^{(0)}=e^{x}\sum_{n=1}A_{n}x^{-n}.
\end{gather}
Inserting this expression into the above differential equation one obtains the recursion relation for the expansion coefficients $A_k$:
\begin{gather}
A_{n+1}=\frac{A_{n}}{6n}(2n+J)(2n-J-1).
\label{recursion}
\end{gather}
The leading term (\ref{I0asim}) gives $A_1=\frac{1}{3}$ and then (\ref{recursion}) determines the whole set of the expansion coefficients.
In this way we obtain for  the ratio $d^2I^{(1)}_J/I^{(0)}_J$ the expression:
\begin{eqnarray}
x\frac{I_{J}^{(1)}}{I_{J}^{(0)}}&=&x-1-\frac{1}{3x}-\frac{5}{9x^{2}}-\frac{37}{27x^{3}}+\left(\frac{1}{6x}+\frac{5}{18x^{2}}+\frac{13}{18x^{3}}\right)J(J+1)\nonumber\\
&&-\frac{1}{54x^{3}}j^{2}(J+1)^{2}+\mathcal{O}(x^{-4}).
\label{truncseri1}
\end{eqnarray}
The convergence in terms of $x$ for the excitation energy may be improved in two steps. First we write the differential equation for 
$I^{(0)}_J$ in a different form:
\begin{equation}
x\left(x\frac{I_{J}^{(1)}}{I_{J}^{(0)}}\right)'+\left(x\frac{I_{J}^{(1)}}{I_{J}^{(0)}}\right)^{2}-\frac{x-1}{2}\left(x\frac{I_{J}^{(1)}}{I_{J}^{(0)}}\right)-\frac{2x^{2}+J(J+1)}{4}=0.
\end{equation}
The derivative $\left(x\frac{I_{J}^{(1)}}{I_{J}^{(0)}}\right)'$ is further calculated by using  (B. 7)(B. 7)(B. 7)(B. 7)(B. 7)(B. 7)(B. 7)) and thus the above equation becomes a second degree algebraic equation for $x\frac{I_{J}^{(1)}}{I_{J}^{(0)}}$. Solving this equation one obtains:
\begin{equation}
x\frac{I_{J}^{(1)}}{I_{J}^{(0)}}=\frac{1}{2}\left[\frac{x-2}{2}+\sqrt{G_{J}}\right],
\end{equation} 
where we used the notation:
\begin{eqnarray}
G_{J}=&\frac{9}{4}x(x-2)+\left(J+\frac{1}{2}\right)^{2}-\frac{4}{9x}\left(3+\frac{10}{x}+\frac{37}{x^{2}}\right)\nonumber\\
&+\frac{2}{3x}\left(1+\frac{10}{3x}+\frac{13}{x^{2}}\right)J(J+1)-\frac{2J^{2}}{9x^{3}}(J+1)^{2}.
\end{eqnarray}
Concerning the near vibrational regime the final expression for energies is obtained in two steps. First one derives the vibrational limit of the k-th derivative:
\begin{eqnarray}
&&\lim_{d\to 0}\left(d^2\frac{I^{(1)}_J}{I^{(0)}_J}\right)^{(k)}=\frac{1}{(2J+3)^k}\left[\frac{J}{2}(\delta_{k,0}+\delta_{k,1})\right.
\nonumber\\
&&\left.+9\frac{(J+1)(J+2)}{2J+5}\left((\delta_{k,2}+9\frac{\delta_{k,3}}{2J+7}\right)\right],~~k=0,1,2,3.
\end{eqnarray}
Then, truncating the Taylor expansion of $x\frac{I^{(1)}_J}{I^{(0)}_J}$, around the point $x=0$, at the third order one obtains:
\begin{eqnarray}
x\frac{I^{(1)}_J}{I^{(0)}_J}&=&\frac{J}{2}+\frac{J}{2(2J+3)}x+\frac{9}{2}\frac{(J+1)(J+2)}{2J+3)^2(2J+5)}x^2\nonumber\\
&+&\frac{27}{2}\frac{(J+1)(J+2)}{(2J+3)^3(2J+5)(2J+7)}x^3+{\cal O}(x^4).
\end{eqnarray}

\end{document}